\def\yyy{}
\def\mcs{{\cal M}_{CS}}
\def\cxH{D}
\input harvmac
%\draftmode
\noblackbox
\def\vx#1{\vec{#1}}
\def\SS#1{^{(#1)}}

\def\zh{\hat{z}}\def\that{{\hat{t}}}
\def\lra{\longrightarrow}

\def\cW{W}\def\cWc{W^{crit}}
\def\tp{\tilde{t}}
\def\hx#1{{\hat{#1}}}

\def\Im{{\rm Im}\, }
\def\Ds{\Delta^*}\def\D{\Delta}
%%%
\lref\Yau{S.~Li, B.~H.~Lian and S.~T.~Yau,
``Picard-Fuchs Equations for Relative Periods and Abel-Jacobi Map for
Calabi-Yau Hypersurfaces,''
arXiv:0910.4215 [math.AG].
%%CITATION = ARXIV:0910.4215;%%
}
\lref\vafaln{C.~Vafa,
 ``Superstrings and topological strings at large N,''
 J.\ Math.\ Phys.\  {\bf 42}, 2798 (2001)
 [arXiv:hep-th/0008142].}
\lref\GMM{
J.~Gomis, F.~Marchesano and D.~Mateos,
``An open string landscape,''
JHEP {\bf 0511}, 021 (2005)
[arXiv:hep-th/0506179].
%%CITATION = JHEPA,0511,021;%%
}
\lref\Ma{
L.~Martucci,
``D-branes on general N = 1 backgrounds: Superpotentials and D-terms,''
JHEP {\bf 0606}, 033 (2006)
[arXiv:hep-th/0602129].
%%CITATION = JHEPA,0606,033;%%
}
\lref\JL{H.~Jockers and J.~Louis,
``The effective action of D7-branes in N = 1 Calabi-Yau orientifolds,''
Nucl.\ Phys.\  B {\bf 705}, 167 (2005)
[arXiv:hep-th/0409098].
%%CITATION = NUPHA,B705,167;%%
}\lref\BBG{M.~Baumgartl, I.~Brunner and M.~R.~Gaberdiel,
``D-brane superpotentials and RG flows on the quintic,''
JHEP {\bf 0707}, 061 (2007)
[arXiv:0704.2666 [hep-th]].}
\lref\KSii{J.~Knapp and E.~Scheidegger,
``Matrix Factorizations, Massey Products and F-Terms for Two-Parameter
Calabi-Yau Hypersurfaces,''
arXiv:0812.2429 [hep-th].}
\lref\sugra{
E.~Cremmer, B.~Julia, J.~Scherk, S.~Ferrara, L.~Girardello and P.~van Nieuwenhuizen, 
``Spontaneous Symmetry Breaking And Higgs Effect In Supergravity Without Cosmological Constant,'' Nucl.\ Phys.\  B {\bf 147}, 105 (1979); %%CITATION = NUPHA,B147,105;%%
E.~Cremmer, S.~Ferrara, L.~Girardello and A.~Van Proeyen,
``Yang-Mills Theories With Local Supersymmetry: Lagrangian, Transformation
Laws And Superhiggs Effect,''
Nucl.\ Phys.\  B {\bf 212}, 413 (1983).  %%CITATION = NUPHA,B212,413;%%
}
\lref\Ltd{W.~Lerche,
``Fayet-Iliopoulos potentials from four-folds,''
JHEP {\bf 9711}, 004 (1997)
[arXiv:hep-th/9709146];
%%CITATION = JHEPA,9711,004;%%
}
\lref\PMff{P.~Mayr,
``Mirror symmetry, N = 1 superpotentials and tensionless strings on
Calabi-Yau four-folds,''
Nucl.\ Phys.\  B {\bf 494}, 489 (1997)
[arXiv:hep-th/9610162].
%%CITATION = NUPHA,B494,489;%%
}
\lref\KSi{J.~Knapp and E.~Scheidegger,
``Towards Open String Mirror Symmetry for One-Parameter Calabi-Yau
Hypersurfaces,''
arXiv:0805.1013 [hep-th].}
\lref\Kiii{
J.~Knapp,
``Deformation Theory of Matrix Factorizations and F-Terms in the Boundary
Changing Sector,''
arXiv:0907.5336 [hep-th].
%%CITATION = ARXIV:0907.5336;%%
}
\lref\GMPh{B.~R.~Greene, D.~R.~Morrison and M.~R.~Plesser,
``Mirror manifolds in higher dimension,''
Commun.\ Math.\ Phys.\  {\bf 173}, 559 (1995)[arXiv:hep-th/9402119].
%%CITATION = CMPHA,173,559;%%
}
\lref\bmn{P.~Berglund and P.~Mayr,
``Non-perturbative superpotentials in F-theory and string duality,''
arXiv:hep-th/0504058.
%%CITATION = HEP-TH/0504058;%%
}
\lref\bmff{P.~Berglund and P.~Mayr,
``Heterotic string/F-theory duality from mirror symmetry,''
Adv.\ Theor.\ Math.\ Phys.\  {\bf 2}, 1307 (1999)
[arXiv:hep-th/9811217].
%%CITATION = 00203,2,1307;%%
}
\lref\LazaroiuNM{
C.~I.~Lazaroiu,
``String field theory and brane superpotentials,''
JHEP {\bf 0110}, 018 (2001)
[arXiv:hep-th/0107162].
}
\lref\AspinwallBS{
P.~S.~Aspinwall and S.~H.~Katz,
``Computation of superpotentials for D-Branes,''
Commun.\ Math.\ Phys.\  {\bf 264}, 227 (2006)
[arXiv:hep-th/0412209].
}
\lref\DouglasFR{
M.~R.~Douglas, S.~Govindarajan, T.~Jayaraman and A.~Tomasiello,
``D-branes on Calabi-Yau manifolds and superpotentials,''
Commun.\ Math.\ Phys.\  {\bf 248}, 85 (2004)
[arXiv:hep-th/0203173].
} 
\lref\Forbes{  B.~Forbes,
``Open string mirror maps from Picard-Fuchs equations on relative
cohomology,''
arXiv:hep-h/0307167.
%%CITATION = HEP-TH/0307167;%%
}
\lref\OOY{H.~Ooguri, Y.~Oz and Z.~Yin,
``D-branes on Calabi-Yau spaces and their mirrors,''
Nucl.\ Phys.\  B {\bf 477}, 407 (1996)
[arXiv:hep-th/9606112].
%%CITATION = NUPHA,B477,407;%%
}
\lref\Kar{M. Karoubi and C. Leruste, ``Algebraic Topology via Differential Geometry'', Cambridge University Press 1987.}
\lref\katzi{S.~Katz, 
``Degenerations of quintic threefolds and their lines'', 
Duke Math. J. 50 (1983), no. 4, 1127.}
\lref\katzii{
S.~Katz,
``Rational Curves On Calabi-Yau Threefolds'', in
Essays on Mirror Manifolds, ed. by S-T Yau, (International Press, 1992);
arXiv:alg-geom/9312009.
%%CITATION = ALG-GEOM/9301006;%%
}
\def\h{\fc{1}{2}}
%%%
\def\abstract#1{
\vskip .5in\vfil\centerline
{\bf Abstract}\penalty1000
{{\smallskip\ifx\answ\bigans\leftskip 2pc \rightskip 2pc
\else\leftskip 5pc \rightskip 5pc\fi
\noindent\abstractfont \baselineskip=12pt
{#1} \smallskip}}
\penalty-1000}
\def\us#1{\underline{#1}}
\def\hth/#1#2#3#4#5#6#7{{\tt hep-th/#1#2#3#4#5#6#7}}
\def\nup#1({Nucl.\ Phys.\ $\us {B#1}$\ (}
\def\plt#1({Phys.\ Lett.\ $\us  {B#1}$\ (}
\def\cmp#1({Comm.\ Math.\ Phys.\ $\us  {#1}$\ (}
\def\prp#1({Phys.\ Rep.\ $\us  {#1}$\ (}
\def\prl#1({Phys.\ Rev.\ Lett.\ $\us  {#1}$\ (}
\def\prv#1({Phys.\ Rev.\ $\us  {#1}$\ (}
\def\mpl#1({Mod.\ Phys.\ Let.\ $\us  {A#1}$\ (}
\def\ijmp#1({Int.\ J.\ Mod.\ Phys.\ $\us{A#1}$\ (}
%%%
\def\br{\hfill\break}\def\noi{\noindent}
\def\cx#1{{\cal #1}}\def\al{\alpha}\def\IP{{\bf P}}\def\IZ{{\bf Z}}\def\IC{{\bf C}}
\def\tx#1{{\tilde{#1}}}\def\bb#1{{\bar{#1}}}
\def\fc#1#2{{#1 \over #2}}\def\p{\partial}
\def\frac#1#2{{#1 \over #2}}\def\p{\partial}
\def\be{\beta}\def\al{\alpha}\def\om{\omega}\def\Om{\Omega}
\def\subsubsec#1{\ \br \noindent {\it #1} \br}
\def\th{\theta}\def\zh{\hx z}

\def\tablecaption#1#2{\kern.75truein\lower0truept\hbox{\vbox{\hsize=5truein\noindent{\bf Table\hskip5truept#1:} #2}}}

\def\-{\hphantom{-}}
\def\Ph{{P_{\cxH}}}
\def\Ga{\Gamma}\def\ga{\gamma}

\def\ux#1{\underline{#1}}
%%%%%%%%%%%%%%%%%%%
\lref\TTstar{S.~Cecotti and C.~Vafa,
``Topological antitopological fusion,''
Nucl.\ Phys.\  B {\bf 367}, 359 (1991).
%%CITATION = NUPHA,B367,359;%%
}
\lref\BCOV{
M.~Bershadsky, S.~Cecotti, H.~Ooguri and C.~Vafa,
``Kodaira-Spencer theory of gravity and exact results for quantum string
amplitudes,''
Commun.\ Math.\ Phys.\  {\bf 165}, 311 (1994)
[arXiv:hep-th/9309140].}
\lref\WaN{
A.~Neitzke and J.~Walcher,
``Background Independence and the Open Topological String Wavefunction,''
arXiv:0709.2390 [hep-th].} 
\lref\CanL{
P.~Candelas, X.~de la Ossa and S.~H.~Katz,
``Mirror symmetry for Calabi-Yau hypersurfaces in weighted P**4 and
extensions of Landau-Ginzburg theory,''
Nucl.\ Phys.\  B {\bf 450}, 267 (1995)
[arXiv:hep-th/9412117].} 
\lref\WLlec{
W.~Lerche,
``Special geometry and mirror symmetry for open string backgrounds with N  =
1 supersymmetry,''
arXiv:hep-th/0312326.
%%CITATION = HEP-TH/0312326;%%
}
\lref\LVW{
W.~Lerche, C.~Vafa and N.~P.~Warner,
``Chiral Rings in N=2 Superconformal Theories,''
Nucl.\ Phys.\  B {\bf 324}, 427 (1989).}
\lref\can{P.~Candelas, X.~C.~De La Ossa, P.~S.~Green and L.~Parkes,
``A pair of Calabi-Yau manifolds as an exactly soluble superconformal
theory,''
Nucl.\ Phys.\  B {\bf 359}, 21 (1991)}
\lref\SYZ{A.~Strominger, S.~T.~Yau and E.~Zaslow,
``Mirror symmetry is T-duality,''
Nucl.\ Phys.\  B {\bf 479}, 243 (1996)
[arXiv:hep-th/9606040].}
\lref\BS{V.~V.~Batyrev and D.~van Straten,
``Generalized hypergeometric functions and rational curves on Calabi-Yau
complete intersections in toric varieties,''
Commun.\ Math.\ Phys.\  {\bf 168}, 493 (1995)
[arXiv:alg-geom/9307010].}
\lref\AHMM{ M.~Alim, M.~Hecht, P.~Mayr and A.~Mertens,
``Mirror Symmetry for Toric Branes on Compact Hypersurfaces,''
arXiv:0901.2937 [hep-th].}
\lref\HaackDI{
M.~Haack, J.~Louis and M.~Marquart,
``Type IIA and heterotic string vacua in D = 2,''
Nucl.\ Phys.\  B {\bf 598}, 30 (2001)
[arXiv:hep-th/0011075].
%%CITATION = NUPHA,B598,30;%%
}
\lref\HerbstJP{
M.~Herbst, C.~I.~Lazaroiu and W.~Lerche,
``Superpotentials, A(infinity) relations and WDVV equations for open
topological strings,''
JHEP {\bf 0502}, 071 (2005)
[arXiv:hep-th/0402110].
}
\lref\LazaroiuRK{
C.~I.~Lazaroiu,
``On the structure of open-closed topological field theory in two
dimensions,''
Nucl.\ Phys.\  B {\bf 603}, 497 (2001)
[arXiv:hep-th/0010269].
} 
\lref\GVW{
S.~Gukov, C.~Vafa and E.~Witten,
``CFT's from Calabi-Yau four-folds,''
Nucl.\ Phys.\  B {\bf 584}, 69 (2000)
[Erratum-ibid.\  B {\bf 608}, 477 (2001)]
[arXiv:hep-th/9906070].
}
\lref\TV{
T.~R.~Taylor and C.~Vafa,
``RR flux on Calabi-Yau and partial supersymmetry breaking,''
Phys.\ Lett.\  B {\bf 474}, 130 (2000)
[arXiv:hep-th/9912152].
}
\lref\PMsp{
P.~Mayr,
``On supersymmetry breaking in string theory and its realization in brane
worlds,''
Nucl.\ Phys.\  B {\bf 593}, 99 (2001)
[arXiv:hep-th/0003198].
}
\lref\BerglundDM{
P.~Berglund and P.~Mayr,
``Non-perturbative superpotentials in F-theory and string duality,''
arXiv:hep-th/0504058.
}
\lref\GovindarajanUY{
S.~Govindarajan and H.~Jockers,
``Effective superpotentials for B-branes in Landau-Ginzburg models,''
JHEP {\bf 0610}, 060 (2006)
[arXiv:hep-th/0608027].
}
\lref\HoriCK{
K.~Hori, A.~Iqbal and C.~Vafa,
``D-branes and mirror symmetry,''
arXiv:hep-th/0005247.
}
\lref\JockersZY{
H.~Jockers and J.~Louis,
``D-terms and F-terms from D7-brane fluxes,''
Nucl.\ Phys.\  B {\bf 718}, 203 (2005)
[arXiv:hep-th/0502059].
}
\lref\KatzGH{
S.~H.~Katz and E.~Sharpe,
``D-branes, open string vertex operators, and Ext groups,''
Adv.\ Theor.\ Math.\ Phys.\  {\bf 6}, 979 (2003)
[arXiv:hep-th/0208104].}  
\lref\LustBD{
D.~L\"ust, P.~Mayr, S.~Reffert and S.~Stieberger,
``F-theory flux, destabilization of orientifolds and soft terms on
D7-branes,''
Nucl.\ Phys.\  B {\bf 732}, 243 (2006)
[arXiv:hep-th/0501139].
}
\lref\SharpeDR{
E.~Sharpe,
``Lectures on D-branes and sheaves,''
arXiv:hep-th/0307245.
}
\lref\WittenZZ{
E.~Witten,
``Mirror manifolds and topological field theory,''
arXiv:hep-th/9112056.
}
%%%%%%%%%%%%%%%%%%%%%%%%%%%%%%
\def\ux#1{{\underline #1}}
\def\IR{{\bf R}}
\def\Res{{\rm Res }}
\def\tha{\vartheta}
\def\thz{\theta}
\def\relform#1#2{{\bf( } #1 , #2{\bf )}}
\def\Si{\Sigma}\def\La{\Lambda}

%%%%%%%%%%%%%%%%
\noi
\def\vr{\vrule height 10pt depth 4pt}
\def\de{\delta}%
\def\mn{\the\secno.\the\subsecno}
\input xymatrix\input xyarrow
\parskip=4pt plus 15pt minus 1pt
\baselineskip=15pt plus 2pt minus 1pt
\lref\BBG{M.~Baumgartl, I.~Brunner and M.~R.~Gaberdiel,
``D-brane superpotentials and RG flows on the quintic,''
JHEP {\bf 0707}, 061 (2007)
[arXiv:0704.2666 [hep-th]].}
\lref\KSii{J.~Knapp and E.~Scheidegger,
``Matrix Factorizations, Massey Products and F-Terms for Two-Parameter
Calabi-Yau Hypersurfaces,''
arXiv:0812.2429 [hep-th].}
\lref\WitCh{E.~Witten,
``Branes and the dynamics of {QCD},
Nucl.\ Phys.\  B {\bf 507}, 658 (1997)
[arXiv:hep-th/9706109].}
\lref\Katz{S.~Kachru, S.~H.~Katz, A.~E.~Lawrence and J.~McGreevy,
``Open string instantons and superpotentials,''
Phys.\ Rev.\  D {\bf 62}, 026001 (2000)
[arXiv:hep-th/9912151]; ``Mirror symmetry for open strings,''
Phys.\ Rev.\  D {\bf 62}, 126005 (2000)
[arXiv:hep-th/0006047].}
\lref\KT{A.~Klemm and S.~Theisen,
``Considerations of one modulus Calabi-Yau compactifications: Picard-Fuchs
equations, K\"ahler potentials and mirror maps,''
Nucl.\ Phys.\  B {\bf 389}, 153 (1993)
[arXiv:hep-th/9205041].}
\lref\CK{D.A. Cox and S. Katz, ``Mirror Symmetry and Algebraic Geometry'',
Mathematical Surveys and Monographs
Vol. 68, American Mathematical Society, 1999.}
\lref\mibo{K. Hori et. al, ``Mirror Symmetry'', Clay Mathematics Monographs v.1., 2003}
\lref\hkty{ S.~Hosono, A.~Klemm, S.~Theisen and S.~T.~Yau,
``Mirror Symmetry, Mirror Map And Applications To Calabi-Yau Hypersurfaces,''
Commun.\ Math.\ Phys.\  {\bf 167}, 301 (1995)
[arXiv:hep-th/9308122].}
\lref\GHKK{T.~W.~Grimm, T.~W.~Ha, A.~Klemm and D.~Klevers,
``The D5-brane effective action and superpotential in N=1
compactifications,''
arXiv:0811.2996 [hep-th].}
\lref\Kont{M.~Kontsevich, ``Homological Algebra of Mirror Symmetry,'' 
Proc. Internat. Congress Math. {\bf 1} (1995) 120-139,
arXiv:alg-geom/9411018.}
\lref\PaWa{R.~Pandharipande, J.~Solomon and J.~Walcher, 
``Disk enumeration on the quintic 3-fold,'' arXiv.org:math/0610901.}
\lref\HL{R.~Harvey and B.~Lawson,
``Calibrated geometries,''
Acta Math.\  {\bf 148}, 47 (1982).}
\lref\mafa{H.~Jockers and W.~Lerche,
``Matrix Factorizations, D-Branes and their Deformations,''
Nucl.\ Phys.\ Proc.\ Suppl.\  {\bf 171}, 196 (2007)
[arXiv:0708.0157 [hep-th]];\br
J.~Knapp,
``D-Branes in Topological String Theory,''
arXiv:0709.2045 [hep-th].}
\lref\VafOS{ C.~Vafa,
``Extending mirror conjecture to Calabi-Yau with bundles,''
arXiv:hep-th/9804131.}
\lref\comp{{\tt Puntos}, http://www.math.ucdavis.edu/~deloera/RECENT\_WORK/puntos2000;\br
{\tt Schubert}, http://math.uib.no/schubert;\br 
{\tt Instanton}, http://www.math.uiuc.edu/~katz/software.html}
\lref\rgkz{I.\ Gel'fand, M.\ Kapranov and A.\ Zelevinsky,
``Hypergeometric functions and toric varieties'', 
Funct. Anal. Appl. {\bf 23} no. 2 (1989), 12.}
\lref\mhvi{C.~Voisin, ``Hodge theory and complex algebraic geometry'',  
Cambridge Studies in Advanced Mathematics 76,77, Cambridge University Press, 2003;\br
C.A.M. Peters and J.H.M. Steenbrink, ``Mixed Hodge Structures'', A Series of Modern Surveys in Mathematics, Vol. 52, Springer, 2008.}
\lref\mhvii{P. Deligne, ``Theori\'e de Hodge I-III'',
Actes de Congr\`es international de Mathematiciens (Nice,1970),
Gauthier-Villars {\bf 1} 425 (1971);
Publ. Math. IHES {\bf 40} 5 (1971);
Publ. Math. IHES {\bf 44} 5 (1974);\br
J. Carlson, M. Green, P. Griffiths and J. Harris,
``Infinitesimal variations of 
Hodge structure (I)'', Compositio Math. {\bf 50} (1983), no. 2-3, 109.}
\lref\mhviii{P.\ Griffiths, ``^<
On the periods of certain rational integrals. I, II,'' 
Ann. of Math. {\bf 90} (1969) 460,496;
``A theorem concerning the differential equations satisfied
by normal functions associated to algebraic cycles'',
Am. J. Math {\bf 101} 96 (1979).}
\lref\WitCS{
E.~Witten,
``Chern-Simons Gauge Theory As A String Theory,''
Prog.\ Math.\  {\bf 133}, 637 (1995)
[arXiv:hep-th/9207094].
%%CITATION = PMTMA,133,637;%%
}
\lref\WittenWDVV{
E.~Witten,
``On the structure of the topological phase of two-dimensional gravity,"
Nucl.\ Phys.\  B {\bf 340}, 281 (1990).
%%CITATION = NUPHA,B340,281;%%
}
\lref\WDVV{
R.~Dijkgraaf, H.~L.~Verlinde and E.~P.~Verlinde,
``Topological Strings In D $<$ 1,''
Nucl.\ Phys.\  B {\bf 352}, 59 (1991).
%%CITATION = NUPHA,B352,59;%%
}
\lref\JockersNG{
H.~Jockers and W.~Lerche,
``Matrix Factorizations, D-Branes and their Deformations,''
Nucl.\ Phys.\ Proc.\ Suppl.\  {\bf 171}, 196 (2007)
[arXiv:0708.0157 [hep-th]].
} 
\lref\WitHS{E.~Witten,
``New Issues In Manifolds Of SU(3) Holonomy,''
Nucl.\ Phys.\  B {\bf 268}, 79 (1986).
%%CITATION = NUPHA,B268,79;%%
}
\lref\Bat{ V.~V.~Batyrev,
``Dual polyhedra and mirror symmetry for Calabi-Yau hypersurfaces in toric
varieties,''
J.\ Alg.\ Geom.\  {\bf 3}, 493 (1994). 
}
\lref\wip{M.~Alim, M.~Hecht, H.~Jockers, P.~Mayr, A.~Mertens, M.~Soroush, in preparation.}
\lref\AV{
M.~Aganagic and C.~Vafa,
``Mirror symmetry, D-branes and counting holomorphic discs,''
arXiv:hep-th/0012041.
%%CITATION = HEP-TH/0012041;%%
}
\lref\AVp{
M.~Aganagic and C.~Vafa, private communication.
}
\lref\glsm{
E.~Witten,
``Phases of N = 2 theories in two dimensions,''
Nucl.\ Phys.\  B {\bf 403}, 159 (1993)
[arXiv:hep-th/9301042];
%%CITATION = NUPHA,B403,159.
}
\lref\glsmii{
P.~S.~Aspinwall, B.~R.~Greene and D.~R.~Morrison,
``Calabi-Yau moduli space, mirror manifolds and spacetime topology  change in
string theory,''
Nucl.\ Phys.\  B {\bf 416}, 414 (1994)
[arXiv:hep-th/9309097].
}
\lref\MW{
D.~R.~Morrison and J.~Walcher,
``D-branes and Normal Functions,''
arXiv:0709.4028 [hep-th].}
\lref\HV{
K.~Hori and C.~Vafa,
``Mirror symmetry,''
arXiv:hep-th/0002222.
%%CITATION = HEP-TH/0002222;%%
}
\lref\Wa{
J.~Walcher,
``Opening mirror symmetry on the quintic,''
Commun.\ Math.\ Phys.\  {\bf 276}, 671 (2007)
[arXiv:hep-th/0605162].
%%CITATION = CMPHA,276,671;%%
}
\lref\JSii{H.~Jockers and M.~Soroush,
``Relative periods and open-string integer invariants for a compact
Calabi-Yau hypersurface,''
arXiv:0904.4674 [hep-th].}
\lref\JS{H.~Jockers and M.~Soroush,
``Effective superpotentials for compact D5-brane Calabi-Yau geometries,''
Commun.\ Math.\ Phys.\  {\bf 290}, 249 (2009)
[arXiv:0808.0761 [hep-th]].
%%CITATION = CMPHA,290,249;%%
}
\lref\LM{ W.~Lerche and P.~Mayr,
``On N = 1 mirror symmetry for open type II strings,''
arXiv:hep-th/0111113.}
\lref\LMW{ W.~Lerche, P.~Mayr and N.~Warner,
``N = 1 special geometry, mixed Hodge variations and toric geometry,''
arXiv:hep-th/0208039; 
``Holomorphic N = 1 special geometry of open-closed type II strings,''
arXiv:hep-th/0207259.}
\lref\PM{P.~Mayr,
``N = 1 mirror symmetry and open/closed string duality,''
Adv.\ Theor.\ Math.\ Phys.\  {\bf 5}, 213 (2002)
[arXiv:hep-th/0108229].}
\lref\OV{
H.~Ooguri and C.~Vafa,
``Knot invariants and topological strings,''
Nucl.\ Phys.\  B {\bf 577}, 419 (2000)
[arXiv:hep-th/9912123].
%%CITATION = NUPHA,B577,419;%%
}
\lref\AKV{M.~Aganagic, A.~Klemm and C.~Vafa,
``Disk instantons, mirror symmetry and the duality web,''
Z.\ Naturforsch.\  A {\bf 57}, 1 (2002)
[arXiv:hep-th/0105045].
%%CITATION = ZNTFA,A57,1;%%
}
\lref\Heho{
M.~Herbst, K.~Hori and D.~Page,
``Phases Of N=2 Theories In 1+1 Dimensions With Boundary,''
arXiv:0803.2045 [hep-th].
%%CITATION = ARXIV:0803.2045;%%
}
\lref\Forbes{  B.~Forbes,
``Open string mirror maps from Picard-Fuchs equations on relative
cohomology,''
arXiv:hep-th/0307167.
%%CITATION = HEP-TH/0307167;%%
}
%%%%%%%%%%%%%%%%%%%%%%%%%%%%%%%%%%%%% %%%%%%%%%%%%%%%%%%%%%%%%%%

\Title{\vbox{
\rightline{\vbox{\baselineskip12pt
\hbox{LMU-ASC 37/09}
\hbox{SU-ITP-09/41}
\vskip -0.8cm
}}}}
{Hints for Off-Shell Mirror Symmetry }
\vskip -0.8cm
\centerline{\titlefont in type II/F-theory Compactifications}
\vskip 0.7cm
\centerline{M. Alim${}^{a,c}$, M. Hecht${}^a$, H. Jockers${}^b$, P. Mayr${}^a$, A. Mertens${}^a$ and M. Soroush${}^b$}
\vskip 0.6cm
\centerline{\it ${}^a$Arnold Sommerfeld Center for Theoretical Physics}
\centerline{\it LMU, Theresienstr. 37, D-80333 Munich, Germany}
\vskip 0.1cm
\centerline{\it ${}^b$Department of Physics, Stanford University}
\centerline{\it Stanford, CA 94305-4060, USA}
\vskip 0.1cm
\centerline{\it ${}^c$ Hausdorff Center for Mathematics}
\centerline{\it Endenicher Allee 62, D-53115 Bonn, Germany}
\vskip 0cm
\abstract{\noi
We perform a Hodge theoretic study of parameter dependent families 
of D-branes on compact Calabi--Yau manifolds in type II and F-theory
compactifications.
Starting from a geometric Gauss-Manin connection for $B$-type branes
we study the integrability and flatness conditions. The $B$-model
geometry defines an interesting ring structure of operators.
For the mirror $A$-model this indicates the existence of an
open-string extension of the so-called $A$-model connection, 
whereas the discovered ring structure should be part of the open-string
$A$-model quantum cohomology.
We obtain predictions for genuine Ooguri-Vafa invariants for
Lagrangian branes on the quintic in $\IP^4$ that pass some
non-trivial consistency checks. 
We discuss the lift of the brane compactifications to F-theory on 
Calabi--Yau four-folds and the effective couplings in the 
effective supergravity action as determined by the $\cx N=1$ special
geometry of the open-closed deformation space.
\vskip20pt
} 
\Date{\vbox{\hbox{\sl {\hskip0.8cm September 2009}}
}}
\goodbreak

\newsec{Introduction}
Much of the success of closed string mirror symmetry relies on the
identification of deformation families of topological $A$- and $B$-models
obtained by perturbing a reference theory by marginal operators. An important
application is the construction of the mirror map and the so-called 
$A$-model connection from the flat sections of the Gauss-Manin connection 
in the $B$-model. Given the understanding of mirror symmetry 
as an equivalence of D-brane categories \Kont\ it seems natural to contemplate
on an open string version of mirror symmetry which identifies
families of open topological $A$- and $B$-models perturbed by both 
open and closed string operators.

There is an immediate puzzle, however, as there is a superpotential 
in the open-string sector computed by a Chern-Simons functional \WitCS.
A non-trivial extension of the mirror map to the 
open string deformation space therefore appears to require some sort of 
off-shell version of mirror symmetry. This somewhat worrisome issue has been
successfully circumvented  
for non-compact toric branes as defined in \AV, which led 
to many important developments in open topological
strings and matrix models, and for compact, but apparently rigid branes,
which have been recently considered in a series of works following ref.~\Wa.

Nevertheless it could be rewarding to pursue the idea of a 
version of mirror symmetry that identifies families of $A$- and $B$-models
perturbed by closed and open string deformations. Such a map  
would likely involve interesting mathematical structures,
such as a non-trivial open-closed chiral ring and an open-string
version of quantum cohomology as its geometrical $A$-model counterpart. Indeed,
open-string deformations are important in 
the definition of enumerative Ooguri-Vafa invariants \OV\ for world-sheets
with boundaries, and it seems natural to ask about an $A$-model 
quantum product specified by these open-string invariants.

In this note we approach the problem from a Hodge theoretic point of
view, using a certain  relative cohomology space introduced 
in refs.~\refs{\LMW,\JS} as a geometric model for the open-closed
deformation space of the $B$-model. 
Starting from the Gauss-Manin connection on this space, we 
derive some predictions on an open string extension of the so-called
$A$-model connection and the associated invariants, an interesting
quantum ring structure of operators and a conjectural metric on the 
open-closed deformation space. We obtain explicit results for an example of a 
family of branes on the quintic in $\IP^4$,
which satisfy some non-trivial consistency checks.
We stress already at this point that our arguments rely mostly on the
geometric Hodge structure for the $B$-model and we 
do not have a proper understanding of the $A$-model 
side and a world-sheet derivation of these results
at the moment.
This would involve, 
amongst others, a better understanding of the (off-shell) bulk-boundary 
ring and its realization in the $A$-model. Instead we work out 
some predictions on these structures from the geometric ansatz on 
the $B$-model side.

The open-closed deformation space $\cx M$ defined in this way
is not a generic K\"ahler manifold as allowed by the effective $\cx N=1$
supergravity, but of a restricted type controlled by the $\cx N=1$ special geometry
of ref.~\LMW. 
This can be explained by the fact that although the derivations in this note 
start from mirror pairs of branes on Calabi--Yau three-folds, there is a 
close connection to Calabi--Yau four-fold compactifications of a type discussed 
in \refs{\PM,\AHMM}. As will be argued below, the ($B$-type version) of
the four-fold compactification defines an F-theory embedding for the
compactification of the $B$-type branes on the four-fold $Z^*$. 
A crucial point is the existence of a weak (de-)coupling limit $g_s\to 0$
in the $B$-brane geometry, which we identify as the mirror of a 
``large base'' limit of the fibration in the mirror four-fold. Amongst others, the F-theory dual determines the superpotential and the K\"ahler potential of 
the $\cx N=1$ effective supergravity theory as well as subleading corrections in 
the string coupling $g_s$, which should be relevant for phenomenological issues.

The organization of this note is as follows. In sect.~2 we introduce
the deformation space $\cx M$ of a certain family of 
$B$-branes on compact Calabi-Yau three-folds.
In sect.~3 we derive a generalized hypergeometric 
system for the Hodge variation on the relevant cohomology space.
In sect.~4 we study the flatness and integrability conditions of the 
Gauss-Manin connection and define the mirror map on the open-closed
deformation space. Sect.~5 contains a case study
for a family of branes on the quintic and a prediction for Ooguri--Vafa invariants.
In sect.~6 we make some remarks from the world-sheet and CFT point of view
together with a discussion on the open-closed chiral ring. In sect.~7 we discuss the effective 
$\cx N=1$ supergravity action for the type IIB/F-theory compactification and identify
the weak coupling limit. We determine the superpotential and propose a K\"ahler metric on the open-closed 
deformation space $\cx M$; some details of the necessary computations are relegated to app.~A. 
Sect.~8 contains a brief summary of the results and our conclusions. 

%%%%
%%%%

\newsec{Geometry and deformation space of the $B$-model}
We start with the definition of the geometrical structure that will be taken as a model for the 
open-closed deformation space $\cx M$, following refs.~\refs{\LMW,\JS,\AHMM}. Let $(Z,Z^*)$ be a mirror pair of Calabi--Yau three-folds and $(L,E)$ a mirror pair of $A/B$-type branes on it. 
On-shell, the classical $A$-type brane geometry is perturbatively defined by a special Lagrangian submanifold $L\in H_3(Z)$ together with a flat bundle on it \WitCS. At the quantum level non-perturbative open worldsheet instantons may couple to the special Lagrangian submanifold $L$. Then an on-shell quantum $A$-type brane arises if the classical geometry is not destabilized by such instanton corrections \Katz.
The mirror $B$-type geometry consists of a holomorphic sheaf $E$ on $Z^*$ describing a D-brane with holomorphic gauge bundle wrapped on an even-dimensional cycle. The concrete realization and application of open string mirror symmetry to this brane geometry, which will be central to all of the following, has been formulated in the pivotal work \AV. More details on the action of mirror symmetry on brane geometries can be found in refs.~\refs{\OOY,\mibo}.

The moduli space of the closed string $B$-model on $Z^*$ is the space $\mcs$ of complex structures, parametrizing the family $\cx Z^*\to \mcs$ of 3-folds with fiber $Z^*(z)$ at $z\in\mcs$. Here $z=\{z_a\}$, $a=1,...,h^{2,1}(Z^*)$ denote some local coordinates on $\mcs$. An important concept in the Hodge theoretic approach to open string mirror symmetry of refs.~\refs{\LMW,\JS,\AHMM} is the definition of an off-shell deformation space $\cx M$, which includes open string deformations. To study the obstruction superpotential on $\cx M$, one first defines $\cx M$ as an {\it unobstructed} deformation space for a relative homology problem and studies the functions $\ux \Pi^\Si:\cx M \to \IC$ defined by integration over the dual cohomology space. In a second step, one adds an obstruction, which can be shown to induce a superpotential on $\cx M$ proportional to a linear combination of these 'relative periods' $\ux \Pi^\Si$.

The unobstructed moduli space for the family of relative cohomology groups can be defined as the moduli space of a {\it holomorphic} family of hypersurfaces embedded into the family $\cx Z^*$ of CY 3-folds \refs{\LMW,\JS}

\eqn\efam{\eqalign{
\ \hskip-1cm i:\hskip2cm \cx D&\hookrightarrow \cx Z^*\cr
\cxH(z,\zh)&\hookrightarrow Z^*(z)
}}
where\foot{Here and in the following we often use a hat to distinguish data associated with the open string sector.}  $\zh=\{\zh_\al\}$ are local coordinates on the moduli space of the embeddings $i:D(z,\zh)\hookrightarrow Z^*(z)$ for fixed complex structure $z$. The total moduli space $\cx M$ of this family is the fibration
\eqn\diafibm{
\xymatrix{
\hx{\cx M}(\zh)\ar[r]&\cx M \ar[d]^\pi\cr
&\cx M_{CS}(z)}
}
where the point $z\in \cx M_{CS}$ on the base specifies the complex structure on the CY 3-fold $Z^*(z)$ and the point $\zh \in \hx{\cx M}$ on the fiber the embedding. In the context of string theory, the moduli $z$ and $\zh$ arise from states in the closed and open string sector, respectively. Note that the fields associated with the fiber and the base of $\cx M$ couple at a different order in string perturbation theory. This will be relevant when defining a metric on $T\cx M$ in sect.~7.

Following \refs{\LMW,\JL,\JS,\AHMM}, we consider functions on the unobstructed deformation space $\cx M$ given by 'period integrals' on the relative cohomology group defined by the brane geometry. The embedding $i:\, \cxH\hookrightarrow Z^*$ defines the space 
$\Om^*(Z^*,\cxH)$ of relative $p$-forms via the exact sequence 
$$
0\lra \Om^*(Z^*,\cxH) \lra \Om^*(Z^*) 
\ {\buildrel i^* \over \lra}\ \Om^*(\cxH)\lra 0 \ .
$$
The associated long exact sequence defines the relative three-form cohomology group 
\eqn\esrc{
H^3(Z^*,\cxH) \ \simeq \ {\rm ker\ } \big( H^3(Z^*)\to H^3(\cxH)\big) 
\oplus {\rm coker\ } \big( H^2(Z^*)\to H^2(\cxH)\big)\ ,} 
which provides the geometric model for 
the space of groundstates of the open-closed
topological $B$-model. 
In a generic\foot{That is $H^1(Z^*) \simeq 0$ and we made the simplifying assumption that $D$ is ample, which is a reasonable condition on the divisor wrapped by a $B$-type brane. The Lefschetz hyperplane theorem then implies $H^1(D)\simeq 0$ and, by Poincar\'e duality, $H^3(D)\simeq 0$.} situation,
the first summand equals $H^3(Z^*)$ and 
represents the closed string sector capturing the 
deformations of the complex structure
of $Z^*$. The relation of the above sheaf cohomology groups considered in \refs{\LMW,\JS} and the Ext groups studied in ref.~\KatzGH\ will be discussed in sect.~6\yyy. 

By eq.\esrc, a closed 
relative three-form $\ux \Phi\in \Om^3(Z^*,\cxH)$, representing an
element of $H^3(Z^*,\cxH)$, can be described by a 
pair $\relform\Phi\phi$, where $\Phi$ is a 3-form on $Z^*$ and $\phi$ a 
2-form on $\cxH$. The differential is $d\ux\Phi=(d\Phi,i^*\Phi-d\phi)$ and 
the equivalence relation $(\Phi,\phi)\sim (\Phi,\phi)+
(d\al,i^*\al-d\be)$ for $\al\in \Om^2(Z^*)$, $\be\in \Om^1(\cxH)$.
The duality pairing between a 3-chain class $\ga_\Si\in H_3(Z^*,\cxH)$
and a relative $p$-form class $[\ux \Phi]$ is given by the integral
\eqn\defrelper{
\int_{\ga_\Si} \ux\Phi=\int_{int(\ga_\Si)}\Phi-\int_{\p\ga_\Si}\phi\ .
}

The fundamental holomorphic objects of the open-closed 
topological $B$-model are particular examples of eq.\defrelper, namely 
the relative period integrals of the holomorphic
$(3,0)$ form $\ux \Om$ on $Z^*$, viewed as the element $(\Om,0)\in H^3(Z^*,\cxH)$, 
over a basis $\{\ga_\Si\}$ of topological 3-chains:
\eqn\defrelper{
\ux\Pi^\Si(z,\zh)=\int_{\ga_\Si} \ux \Om,\qquad \ga_\Si \in H_3(Z^*,\cxH) \ .
}

The cohomology group $H^3(Z^*,D)$ is constant over $\cx M$, but the Hodge decomposition $F^pH^3(Z^*,D)$ and the direction of the (3,0) form $\ux \Om$ varies with the moduli. The period integrals $\ux\Pi^\Si(z,\zh)$ thus define a set of moduli dependent local functions on $\cx M$. Despite the fact, that there is not yet a superpotential on $\cx M$, these functions should have an important physical meaning in the unobstructed theory as well. In sect.~7 we argue that they define a K\"ahler metric on $\cx M$ and thus determine the kinetic terms of the bulk and brane moduli in the effective action.

Further details on the relation between relative cohomology and open-closed deformation spaces can be found in refs.~\refs{\LMW,\WLlec,\JL,\JS,\Forbes,\GHKK}. For the mathematical background, see e.g. refs.~\refs{\Kar,\mhvi}.\foot{After publication of this work, a thorough mathematical justification and generalization of the methods of refs.\refs{\LMW,\JS,\AHMM} appeared in ref.~\Yau, with some overlap with sect.~3 below.}

\subsubsec{Obstructed deformation problem}
The physical meaning of the period integrals is altered after adding an additional lower-dimensional brane charge on a 2-cycle, which induces an obstruction on $\cx M$. From a physics point of view this perturbation may be realized by either adding an additional brane on a 2-cycle in $\cxH$ or by switching on a 2-form gauge flux on the original brane on $\cxH$. A world-sheet derivation of the obstruction from the relevant Ext groups in the open string CFT will be  given in sect.~6. 

In the Hodge theoretic approach of refs.~\refs{\LMW,\JS,\AHMM}, the superpotential on $\cx M$ in the obstructed theory is given by a certain linear combination of the relative periods \defrelper\ of the unobstructed theory, as reviewed below. This is similar to the case of closed string flux compactifications, where the flux superpotential on the space $\mcs$ of complex structures can be computed in the unobstructed theory with $\mcs$ as a true moduli space \refs{\vafaln,\TV,\PMsp}.

Let $C_i$ denote the irreducible components of the 2-cycle carrying the additional brane charge and $C=\sum_i C_i$ their sum. If $[C]=0$ as a class in $H_2(Z^*)$, there exists a 3-chain $\Ga$ in the sheaf cohomology group \esrc, with $\p \Ga=C$. In particular,  the choice of the brane cycle $C$ restricts the relevant co-homology to the subspace 
\eqn\ssrc{
H_3(Z^*,D)\ \longrightarrow \ H_3(Z^*,\sum_i C_i)\ .
}
The open-closed string superpotential $W(z,\zh)$ on $\cx M$ for this brane configuration is computed by a relative period integral $\ux\Pi(z,\zh)$ on this subspace \refs{\LMW,\JS,\AHMM}.

It was argued in \WitCh, that a superpotential, that has the correct critical points to describe a supersymmetric brane on $C$, is given by the chain integral 
\eqn\defci{
\cx T=\int_{\ga(C)} \Om \qquad \p{\ga(C)}\neq 0\ .}
This expression was later derived from a dimensional reduction of the holomorphic Chern-Simons functional of ref.~\WitCS\ in refs.~\refs{\Katz,\AV}.\foot{More precisely, the chain integral gives the tension $\cx T$ of a domain wall realized by a brane wrapped on the 3-chain $\ga(C)$.}

As it stands, eq.\defci\ can be viewed either as a definition in absolute cohomology, or in relative cohomology, replacing $\Om\to (\Om,0)$ and including the explicit boundary term in eq.\defrelper. The difference is important only off-shell and in this way the relative cohomology ansatz of refs.\refs{\LMW,\JS,\AHMM}, building on the results of \AV, can be viewed as a particular proposal for an off-shell definition of the superpotential.

In absolute cohomology, the integral \defci\ is a priori ill-defined because of 
non-vanishing boundary contributions from exact forms, which do not respect 
the equivalence relation $[\Om]=[\Om+d\om]$. To obtain a well-defined pairing one may restrict homology to chains with boundary $\p\ga$ a holomorphic curve and cohomology to sections of the Hodge subspace $F^2H^3=H^{3,0} \oplus H^{2,1}$ \mhvi.%
\foot{The potentially ambiguous boundary terms then vanish as
$\int_\ga \Om+ d \om=\int_\ga \Om +\int_{\p \ga}\om=\int_\ga \Om$ for $\om$ a (2,0) form 
and $\p \ga$ a 2-cycle of type $(1,1)$.} This is the normal function point of view taken in refs.~\refs{\Wa,\MW}. Since the curve $C=\p \ga$ being holomorphic corresponds to a critical point $d\cW=0$ of the superpotential with respect to the open string moduli \WitCh, continuous open-string deformations are excluded from the beginning and one obtains the critical value $\cWc(z)$ of the superpotential as a function of the closed-string deformations $z$, only. The dependence of the critical superpotential $W_{crit}(z)$ on the closed string moduli $z$ is still a highly interesting quantity and at the center of the works \refs{\Wa,\MW} on open string mirror symmetry, which gave the first computation of disc instantons in compact CY 3-fold from mirror symmetry. The dependence of the superpotential on open string deformations $\zh$ is not captured by this definition.

In the relative cohomology ansatz of refs.~\refs{\LMW,\JS,\AHMM}, the pairing \defci\ is well-defined in cohomology also away from the critical points as a consequence of enlarging the co-homology spaces as in \esrc. The extra contribution to $H^3(Z^*,\cxH)$ from the second factor in \esrc\ describe additional degrees of freedom in the brane sector.  According to this proposal, the relative periods $\ux\Pi(z,\zh)$ on the subspace $H^3(Z^*,C)$ describe  the 'off-shell' superpotential $\cW(z,\zh)$ depending on brane deformations $\zh$. For consistency, $\cW(z,\zh)$ should reduce to the critical superpotential $\cWc(z)$ at the critical points. This has been verified for particular examples in refs.\refs{\JS,\AHMM}.

Although we eventually end up with studying the periods on the restricted subspace $H^3(Z^*,C)$ in \ssrc\ for a fixed brane charge $C$, the introduction of the larger relative cohomology space $H_3(Z^*,D)$ was not redundant, even for fixed choice of obstruction brane $C$, as it was crucial for the definition of the finite-dimensional off-shell deformation space $\cx M$, on which the obstruction superpotential can be defined. The off-shell deformation space for a brane on $C$ is generically infinite-dimensional, with most of the deformations representing heavy fields in space-time that should be integrated out. To define an effective superpotential we have to pick an appropriate set of 'light' fields and integrate out infinitely many others.

The ansatz of refs.~\refs{\LMW,\JS,\AHMM} to define $\cx M$ by perturbing the unobstructed moduli space of a family $\cx D$ of hypersurfaces is thus not a circuitry, but rather a systematic way to define a finite-dimensional deformation space with parametric small obstruction, together with a local coordinate patch, on which a meaningful off-shell superpotential can be defined. As $C$ can be embedded in different families of hyperplanes, the parametrization of the deformation space depends on the choice of the family $\cx D$ and this corresponds to a different choice of light fields for the effective superpotential.\foot{One could always combine these 'different' families into a single larger family at the cost of increasing the dimension of the deformation space $\cx M$.} Each choice covers only a certain patch of the off-shell deformation space and there will be many choices to parametrize the same physics and mathematics near a critical locus by slightly different relative cohomology groups. This choice of a set of light fields is inherent to the use of effective actions and should not be confused with an ambiguity in the definition.\foot{An attempt to reformulate the relative cohomology  approach of refs.~\refs{\LMW,\JS,\AHMM} by using the excision theorem, as contemplated on in ref.~\GHKK, is thus likely to produce just another parametrization corresponding to a slightly different choice of light fields, rather then a distinct description.} 

In the context of open string mirror symmetry, the most interesting aspect of the deformation spaces $\cx M$ constructed in this way is the presence of 'almost flat' directions in the open string sector, which lead to the characteristic $A$-model instanton expansion of the superpotential, as will be shown in in sect.~5. The result passes some non-trivial consistency checks which provides some evidence in favor of this definition of off-shell string mirror symmetry.\foot{See also refs.~\refs{\AHMM,\JSii} for additional examples and arguments.}  On the other hand, for general massive deformations, one would not expect the simple notions of flatness and an integral instanton expansion observed in this paper. 

The most general superpotential captured by the relative cohomology group $H^3(Z^*,C)$ includes also a non-trivial closed string flux on $H^3(Z^*)$ and the two contributions can be combined in the general linear combination of relative period integrals \refs{\LMW,\JS}

\eqn\supi{\eqalign{
\cW_{\cx N=1}(z,\zh)=\sum_{\ga_\Si \in H^3(Z^*,\cxH)}\ux{N}_\Si\,\ux\Pi^\Si(z,\zh)\ =\cW_{closed}(z)+\cW_{open}(z,\zh),
}}
where the contributions from the closed and open string sector are, with  $\ux N_\Si :=(N_\Si,\hx N_\Si)$,
$$
\cW_{closed}(z)=\sum_{\ga_\Si,\p \ga_\Si=0} N_\Si\,\ux\Pi^\Si(z),\qquad 
\cW_{open}(z,\zh)=\sum_{\ga_\Si,\p \ga_\Si\neq 0} \hx N_\Si\,\ux\Pi^\Si(z)\ .
$$
This is the superpotential for a four-dimensional $\cx N=1$ supersymmetric string compactification with $N_\Si$ and $\hat{N}_{\Si}$ quanta of background ``fluxes'' in the closed and open string sector, respectively.
The first term  $\cW_{closed}(z)$ is proportional to the periods over {\it cycles} 
$\ga_\Si\in H^3(Z^*)$ and represents the closed string ``flux'' 
superpotential for $N_\Si$ ``flux'' quanta \refs{\GVW,\TV,\PMsp}. 
The second term captures the chain integrals \defci. Note that there are contributions to the superpotential from different orders in the string coupling and the instanton expansion of the mirror $A$-model will involve {\it sphere and disc instantons} at the same time.\foot{This links to the open-closed string duality to Calabi--Yau 4-folds of refs.~\refs{\PM,\AHMM}; see also \ref\AganagicJQ{
M.~Aganagic and C.~Beem,
``The Geometry of D-Brane Superpotentials,''
arXiv:0909.2245 [hep-th].
%%CITATION = ARXIV:0909.2245;%%
} for a recent discussion.}

There are two important points missing in the above discussion, which will be further studied in the following. One is the selection of the proper homology element $\ga(C)$ that computes the superpotential, given a 2-cycle $C$ representing the lower-dimensional brane charge. The other one is the mirror map, which allows to extract a prediction for the disc and sphere instanton expansion for the $A$-model, starting from the result obtained from the relative periods of the $B$-model. The additional information needed to answer these questions comes from the variation of mixed Hodge structure on the Hodge bundle with fiber the relative cohomology group $H^3(Z^*,D)$. The Hodge filtration defines a grading by Hodge degree $p$ of the cohomology space at each point $(z,\zh)$. In closed string mirror symmetry, restricting to $H^3(Z^*)\subset H^3(Z^*,D)$, this grading is identified with the $U(1)$ charge of the chiral ring elements in the conformal field theory on the string world-sheet. A similar interpretation in terms of an open-closed chiral ring has been proposed in refs.~\refs{\LMW,\JS}. The upshot of this extra structure is, that there are {\it two} relevant relative period integrals associated with the brane charge $C$, distinguished by the grading, such that one gives the mirror map to the $A$-model, and the other one the superpotential \refs{\LMW,\JS,\AHMM}. 

In the following sections we thus turn to a detailed study of the variation of the mixed Hodge 
structure on the relative cohomology group $H^3(Z^*,\cxH)$, which we take 
as a geometric model for the variation of the states of the open-closed 
$B$-model over the deformation space $\cx M$. In sect.~3 we derive a set of differential equations, whose solutions determine a basis for the periods $\ux\Pi^\Si(z,\zh)$ on $\cx M$ in terms of generalized hypergeometric functions. In sect.~4 we study the mixed Hodge variation on the relative cohomology bundle, which leads to the selection of the proper functions for the mirror map and the superpotential.
%%%%
%%%%

\newsec{Generalized hypergeometric systems for the $B$-model}
In the first step we derive a generalized hypergeometric system of 
differential operators for the deformation problem defined above, in the 
concrete framework of toric branes on toric CY hypersurfaces
defined in ref.~\AV\ and further scrutinized in \refs{\AHMM,\JSii}. 
The result is a system of differential equations acting on the relative 
cohomology space and its periods, whose associated Gauss-Manin system
and solutions  will be studied in the next
section. The result is summarized in eq. (3.21\yyy); %\gkz
the reader who is not interested in the derivation may safely 
skip this section.

To 
avoid lengthy repetitions, we refer 
to refs.~\refs{\AV,\AHMM} for the definitions of the family of toric branes 
in compact toric hypersurfaces, 
to refs.~\refs{\CK,\mibo} for background material on mirror symmetry 
and toric geometry and to refs.~\refs{\BS,\hkty,\Bat} 
for generalized hypergeometric systems
for the closed-string case. The notation is as 
follows : $\D$ is a reflexive polyhedron in 
$\IR^5$ defined as the convex hull of 
$p$ integral vertices $\nu_i\subset \IZ^5\subset\IR^5$ lying in a hyperplane
of distance one to the origin.%
\foot{We use the standard convention, identify the interior point $\nu_0$ of $\D$
with the origin, and specify the vertices by four components 
$\nu_{i,k},\ k=1,...,4$, i.e. $\nu_0=(0,0,0,0)$; see refs.~\refs{\Bat,\hkty,\CanL} 
for more details.}
$W=P_{\Sigma(\D)}$ is the toric variety with fan $\Sigma(\D)$
defined by the set of cones over the faces of $\D$. $\Ds$ is the 
dual polyhedron and $W^*$ the toric variety obtained from $\Sigma(\Ds)$.
The mirror pair of toric hypersurfaces in $(W,W^*)$ is denoted by $(Z,Z^*)$.

\subsec{Generalized Hypergeometric systems for Calabi-Yau three-folds}
The $p$ (relevant) integral points of $\D$ determine the hypersurface $Z^*\subset W^*$
as the vanishing locus of the equation 
$$
P(Z^*)=\sum_{i=0}^{p-1} a_iy_i=\sum_{\nu_i\in \D}a_iX^{\nu_i}
$$
where $a_i$ are complex parameters,  
$y_i$ are certain homogeneous coordinates \HV\ on $W^*$, 
$X_k$, $k=1,...,4$ are inhomogeneous coordinates 
on an open torus $(\IC^*)^4\subset W^*$ and $X^{\nu_i}:=\prod_k X_k^{\nu_{i,k}}$ \Bat.
The integral points 
$\nu_i$ and the homogeneous coordinates $y_i$ fulfill $h^{1,1}(Z)=h^{2,1}(Z^*)$ relations
\eqn\elinr{
\sum_{i=0}^{p-1} 
l_i^a\nu_i=0\ , \hskip2cm \prod_{i=0}^{p-1} y_i^{l_i^a}=1\ , 
\qquad a=1,...,h^{1,1}(Z).
}
The $p$-dimensional integral vectors $l^a$ specify the 
charges of the matter fields of the gauged linear sigma model (GLSM)
associated with $Z$ \glsm. The 
index 0 refers to the special field $p$ of negative charge which enters linearly
in the two-dimensional GLSM superpotential.

An important datum for the $B$-model on the mirror manifold $Z^*$ are the 
period integrals of the holomorphic (3,0) form $\Om$.
The fundamental period integral on $Z^*$ can be defined 
as \BS
\eqn\deffp{
\Pi(a_i)= \fc{1}{(2\pi i)^4}\int_{|X_k|=1} \fc{1}{P(Z^*)}\ {\prod_{k=1}^4}\fc{dX_k}{X_k}\ .
}
As noted in refs.~\refs{\BS,\Bat}, the period integral is annihilated by a system 
of differential
operators of the so-called GKZ hypergeometric type \rgkz
\eqn\defgkzi{\eqalign{
\cx L(l) &= \prod_{l_i>0}\big(\fc{\p}{\p{a_i}}\big)^{l_i}-\prod_{l_i<0}\big(\fc{\p}{\p{a_i}}\big)^{-l_i},\qquad  l\in K,\cr
\cx Z_k &= \sum_{i=0}^{p-1}\nu_{i,k}\tha_i\ ,k=1,...,4; \qquad \cx Z_0=\sum_{i=0}^{p-1}\tha_i+1\ .
}}
where $\tha_i=a_i\p_{a_i}$ and $K$ denotes the set of integral linear combinations
of the charge vectors $l^a$. 
The differential equations $\cx L(l)\, \Pi(a_i)=0$ follow straightforwardly 
from the definition \deffp.
The equations $\cx Z_k \, \Pi(a_i)=0$ 
express the invariance of the period integral 
under the torus action and imply that 
the periods depend, up to normalization,
only on special combinations of the parameters $a_i$, $\Pi(a_i) \sim \Pi(z_a)$, where 
\eqn\defmod{
z_a=(-)^{l^a_0} \prod_i a_i^{l^a_i}\ ,
}
define $h^{2,1}(Z^*)=h^{1,1}(Z)$ 
local coordinates on the complex structure moduli space of $Z^*$ \hkty.

\subsec{Extended hypergeometric systems for relative periods}
We proceed with the derivation of a GKZ hypergeometric
system which annihilates the relative period integrals \defrelper\ 
on the relative cohomology group. The definition of the (union of) 
hypersurfaces $\cxH$ cannot preserve all torus symmetries. 
Instead (some of) the torus actions move 
the position of the branes.\foot{This is what one would expect intuitively from 
the formulation of mirror symmetry as $T$ duality \SYZ.} As a consequence, 
the relative periods are no longer annihilated by all the operators 
$\cx Z_k$ and depend on additional parameters specifying the geometry 
of $\cxH$.  The differential equations \defgkzi\ for the period integrals imply 
on the level of forms
$$\eqalign{
\cx L(l)\ \Om &= d\om(l)\ , \cr
\cx Z_k\ \Om &=d\om_k \ .
}$$
The exact terms on the r.h.s. contribute only to integrals 
over 3-chains $\hx \gamma\in H_3(Z^*,\cxH)$
with non-trivial boundaries $\p \hx \ga$. The modification of the 
differential equations for the relative periods can be computed from 
these boundary terms. 

To 
keep the discussion simple we derive the differential operators for 
the relative periods on the mirror of the quintic and present general 
formulae at the end of this section. The integral points of the polyhedron $\Delta$ are
\eqn\defvertsb{\eqalign{
\nu_0 &= ( 0 , 0 , 0 , 0 ) \ , \quad
\nu_1 = ( 1 , 0 , 0 , 0 ) \ ,  \quad
\nu_2 = ( 0 , 1 , 0 , 0 ) \ ,  \quad
\nu_3 = ( 0 , 0 , 1 , 0 ) \ , \cr 
\nu_4 &= ( 0 , 0 , 0 , 1 ) \ , \quad
\nu_5 = ( -1 , -1 , -1 , -1 ) \ , }}
leading to the defining polynomial for the mirror quintic:
$$
P(Z^*)=a_0+a_1X_1+a_2X_2+a_3X_3+a_4X_4+a_5(X_1X_2X_3X_4)^{-1}\ .
$$
The holomorphic (3,0) form on the mirror quintic can be explicitly represented as
the residuum 
\eqn\ResOm{
\Om = \Res_P\ \fc{a_0}{P}\prod_k\fc{dX_k}{X_k}\ , }
at $P=0$ \BS. Here we have changed the normalization with respect to eq.~\deffp\
to the standard convention $\Pi(a_i)\to a_0\Pi(a_i)$. 
The GLSM for the quintic is specified by one charge vector $l^1 = (-5,1,1,1,1,1)$ which defines one differential operator  $\cx L_1=\cx L(l^1)$ annihilating the periods on the mirror. For this operator we do not get an exact term, and we find
\eqn\deflone{
\left(\prod_{i=1}^5\tha_i+z \prod_{i=1}^5(\tha_0-i)\right)\ \Om=0 \ .
}
Furthermore the operators $\cx Z_k$ give rise to the relations 
\eqn\defreli{\sum_{i=0}^5\tha_i\  \Om=0,
\qquad
(\tha_i-\tha_5)\  \Om =d \om_i\ ,\quad i=1,...,4 ,
}
with 
\eqn\exact{\om_i=(-)^{i+1}\Res_P \ \fc{a_0}{P}\prod_{j=1\atop j\neq i}^{4}\fc{dX_j}{X_j}\ .}
Equation \defreli\ expresses the torus 
invariance of the period integrals in absolute cohomology and 
implies that the integrals depend only on the single invariant 
complex modulus $z_1$ defined as in eq.~\defmod. 
In relative cohomology, the exact terms on the r.h.s. descend to 
non-trivial 2-forms on $\cxH$ by the equivalence relation 
\eqn\relcoh{
H^3(Z^*,\cxH) \, \ni \, \relform\Xi\xi \sim \relform{\Xi+d\al}{\xi+i^*\al-d\be}\ ,
}
where $i:\cxH \hookrightarrow Z^*$ is the embedding. The exact pieces
in \defreli\ may give rise to boundary terms that break the torus symmetry 
and introduce an additional dependence on moduli $\zh_\al$ 
associated with the geometry of the embedding of $\cxH$.

To proceed we need to specify the family of hypersurfaces $\cxH$. 
As in refs.~\refs{\JS,\AHMM} we consider a simple 1-parameter family
$\cxH_1$ of hypersurfaces defined by a linear equation, which 
can be put into the standard form
\eqn\defhone{ \cxH_1:\ Q =  1 + X_1=0\ , }
by a coordinate transformation on $X_1$.\foot{At first glance
it seems that we have chosen a rigid hypersurface $\cxH_1$. However, 
as we vary the embedding of the
Calabi-Yau manifold $Z^*$ in the ambient toric space, we effectively change the
hypersurface $\cxH_1$ in in the Calabi-Yau manifold $Z^*$.}
In order to determine the preserved torus symmetries we examine the 
boundary contributions~\defreli\ with respect to the hypersurfaces $\cxH_1$
by evaluating the pullbacks of the two forms~\exact. One finds 
$i^*\om_k = 0$ for $k=2,3,4$ and 
\eqn\deftomi{ i^*\om_1=\Res_{\Ph}\fc{a_0}{\Ph}\prod_{i=2}^4\fc{dX_i}{X_i}\ = \Res_{\!P,Q}\ {a_0X_1\over P\,Q}\prod_{k=1}^4\fc{dX_k}{X_k}\ ,}
where $
\Ph=P(Z^*)|_{Q=0}=(a_0-a_1)+a_2X_2+a_3 X_3+a_4 X_4-a_5(X_2X_3X_4)^{-1}$.
In the second equation above, we represented the pull-back as a 
double residue in $P=0$ and $Q=0$ in the ambient space as in 
ref.~\JS, which allows for a direct evaluation of periods
as in ref.~\JSii\ and is convenient also for higher degree hypersurfaces.

In the presence of the hypersurface $\cxH_1$ the torus action
$X_1\to \lambda X_1$ with $\lambda\in\IC^*$ generated by the operator
$\cx Z_1$ is broken, whereas the remaining torus symmetries, associated to
the operators $\cx Z_0, \cx Z_2, \cx Z_3, \cx Z_4$, are preserved.\foot{Alternatively one may express the independence of the periods on extra parameters in \defhone\
by a modified differential operator $\tx Z_1$, see also the discussion below 
eq. (3.19)\yyy.}  From these
four differential operators for the six parameters $a_i$ it follows that the family
$\cxH_1\subset Z^*$ depends precisely on the two moduli 
\eqn\defmodhone{ z=-\fc{a_1a_2a_3a_4a_5}{a_0^5},\qquad \hx z=-\fc{a_1}{a_0}\ ,}
together with the logarithmic derivatives
\eqn\defdermod{ \theta:=z\,\partial_z = \tha_5 \ , \qquad 
\hx\theta:=\zh\,\partial_\zh =\cx Z_1= \tha_1 - \tha_5 \ . }
Here $z$ is the complex structure modulus of $Z^*$ and $\zh$ is the open-string
position modulus parametrizing the position of the brane. This analysis justifies
in retrospect that the defined family of hypersurfaces \defhone\ indeed depends on a single
open-string modulus.

The next task is to determine the differential operators $\cx L$ of the extended GKZ
hypergeometric system associated to the periods of the relative (3,0) form representative
\eqn\pair{
\ux\Om:=\relform\Om0 = \left( \Res_P {a_0\over P}\prod_k\fc{dX_k}{X_k}\ , 0\right) \ . }
Due to eq.~\deflone\ the operator $\cx L_1$ annihilates the three form $\Om$ even on
the level of forms (and not just on the level of the absolute three-form
cohomology), and therefore the operator $\cx L_1$ annihilates also the relative three form
$\Om$. In the extended relative cohomology setting, however, there is an additional differential
operator $\cx L_2$ governing the functional dependence of the exact piece $d\om_1$. It is
straightforward to check that the (2,0) form $i^*\om_1$ of the hypersurface $\cxH_1$
defined in eq.~\deftomi\ obeys
$$ \left( \tha_1 - \zh (\tha_0-1) \right) i^*\om_1 = 0 \ . $$
Together with the relation
\eqn\defex{\thz_2\ux\Om=\cx Z_1\relform\Om0 = \relform{d\om_1}0 \sim \relform0{-i^*\om_1} \ ,}
associated to the broken torus action, we determine the second operator $\cx L_2$ to be
$$ \cx L_2 = (\tha_1 - \zh (\tha_0 -1))(\tha_1-\tha_5) \ .$$
Thus in summary we find that the relative (3,0)-form cohomology class, represented by the
relative three-form $\ux\Om$, is annihilated by the six differential operators
\eqn\extGKZ{\eqalign{ \cx L_a \ux\Om &\sim 0 \ , \quad a=1,2 \ , \cr
\cx Z_k \ux\Om &\sim 0 \ , \quad k=0,2,3,4 \ . }}
With the help of the differential operators $\cx Z_k, k=0,2,3,4\,,$ it is straightforward to derive
the two extended GKZ operators $\cx L_a, a=1,2\,,$ in terms of the moduli $z$ and $\zh$ and
their logarithmic derivatives $\theta$ and $\hx\theta$
\eqn\GKZ{\eqalign{
\cx L_1 &= (\theta+\hx\theta)\thz^4+z\prod_{i=1}^5 (-5\theta-\hx\theta-i)
=:\cx L_1^{bulk}+\cx L_1^{bdry}\hx\theta\ \ , \cr
\cx L_2 &= \big((\theta+\hx\theta)-\zh(-5\theta-\hx\theta-1)\big)\hx\theta
=:\cx L_2^{bdry}\hx\theta \ . }}
Here $\cx L_1^{bulk}=\theta^5+z\prod_{i=1}^5(-5\theta-i)$ is the 
$\hx\theta$-independent GKZ operator of the quintic and
the operators $\cx L_a^{bdry}$ are always accompanied
by at least one derivative $\hx\theta$ and thus are only sensitive to 
boundary contributions
\eqn\bdrycont{
0 \,=\, \cx L_a \int_{\ga_\Si} \ux\Om \,=\,\cx L_a^{bulk} \int_{int(\ga_\Si)}\Om
- \cx L_a^{bdry} \int_{\p\ga_\Si} i^*\om_1 \ , \quad  a = 1, 2 \ ,}
with $\cx L_2^{bulk}\equiv 0$.

The significance of the interplay of the operator $\hx\theta$ with the operator
$\cx L_a^{bdry}$ is reflected in eq.~\defex.  We observe that the (linear combinations) of
boundary operators $\cx L_a^{bdry}$, which are not accompanied with a non-vanishing bulk operator
$\cx L_a^{bulk}$, become the GKZ operators of the periods localizing on the hypersurface $\cxH_1$.
As noted in ref.~\AHMM, the Hodge variation on the hypersurface $\Ph=0$ is isomorphic to that of the mirror of the quartic K3~surface
and $i^*\om_1$ is a representative for the holomorphic (2,0) form. The associated K3~periods
$\int i^*\om_1$ are precisely annihilated by these GKZ operators arising
form the boundary sector.

Thus we have obtained two differential operators $\cx L_a,\ a=1,2\,,$  
in eq.~\GKZ\ annihilating the relative periods. These operators 
can be rewritten in a concise form by realizing that they represent the
differential operators $\cx L(l)$ for a different GKZ system of the type \defgkzi\
specified by the two linear relations 
\eqn\elti{
\tx l^1=(-5;1,1,1,1,1,0,0),\qquad \tx l^2=(-1;1,0,0,0,0,1,-1)\ ,
}
together with the five-dimensional enhanced toric polyhedron $\tx\Delta$ with the integral vertices \AHMM
\eqn\defverts{\eqalign{
\tx\nu_0 &= ( 0 , 0 , 0 , 0 , 0 ) \ , \quad
\tx\nu_1 = ( 1 , 0 , 0 , 0 , 0 ) \ ,  \quad
\tx\nu_2 = ( 0 , 1 , 0 , 0 , 0 ) \ ,  \quad
\tx\nu_3 = ( 0 , 0 , 1 , 0 , 0 ) \ , \cr 
\tx\nu_4 &= ( 0 , 0 , 0 , 1 , 0 ) \ , \quad
\tx\nu_5 = ( -1 , -1 , -1 , -1 , 0 ) \ ,  \quad
\tx\nu_6 = ( 0 , 0 , 0 , 0 , 1 ) \ ,  \quad
\tx\nu_7 = ( 1 , 0 , 0 , 0 , 1 ) \ . }}
This polyhedron defines a CY four-fold $X^*$, 
which is dual to the brane compactification in the sense of ref.~\PM.  
A systematic construction, which associates a compact dual F-theory four-fold to a mirror pair of toric branes defined as in ref.~\AV,
is relegated to sect.~7. 

Note that the enhanced polyhedron $\tx\Delta$ gives rise to six operators $\tx{\cx Z}_k, k=0,\ldots,5\,,$
with ${\tx{\cx Z}}_k = \cx Z_k$ for $k=0,2,3,4$. The two additional operators
${\tx{\cx Z}}_1$ and ${\tx{\cx Z}}_5$ guarantee that the functional
dependence on the local coordinates $\tx z_a(\tx l^a)$ defined by the general relation \defmod\ also coincide
with the moduli of the relative cohomology problem defined in eq.~\defmodhone. 
Moreover, the GKZ operators $\tx{\cx L}_a=\cx L(\tx l^a)$ obtained from the general expression
\eqn\gkz{
{\cx L}(l)=
\prod_{k=1}^{l_0}(\tha_0-k)\prod_{l_i>0}\prod_{k=0}^{l_i-1}(\tha_i-k)
-(-1)^{l_0} z_a \prod_{k=1}^{-l_0}(\tha_0-k)\prod_{l_i<0}\prod_{k=0}^{-l_i-1}(\tha_i-k)
}
coincide with the two operators $\cx L_a$ of the relative
cohomology problem in the local coordinates defined by \defmod.
$$
\cx L_a(z,\zh)=\tx{\cx L}_a(\tx z_1=z,\tx z_2=\zh)\ .
$$

In fact one can show that all differential operators 
${\cx L}(l)$ for $l$ a linear combination of $\tx l^1,\tx l^2$ also annihilate
the relative periods.\foot{For the quintic the additional operators are of the
form $\cx L^{bdry} \thz_2$, where $\cx L^{bdry} i^*\om_1=0$ modulo exact 2-forms on $\cxH_1$.}
The coincidence of the system of differential operators for the periods 
on the relative cohomology group $H^3(Z^*,\cxH)$ and the GKZ 
system for the dual four-folds constructed by the method of ref.~\AHMM\
holds more generally for relative cohomology
groups associated with the class of 
toric branes on toric hypersurfaces defined in ref.~\AV. 

\newsec{Gauss-Manin connection and integrability conditions}
The Picard-Fuchs system for the relative periods derived in the previous section
captures the variation of Hodge structure on the relative cohomology group 
$H^3(Z^*,\cxH)$. In this section we work out some relations 
and predictions for mirror symmetry for a family of $A$- and $B$-branes
from the associated Gauss-Manin connection.

\subsec{Gauss-Manin connection on the open-closed deformation space $\cx M$}
Geometrically we can view $H^3(Z^*,\cxH)$ as the fiber of a 
complex vector bundle over 
the open-closed deformation space $\cx M$.
As the relative cohomology group\foot{Letters without arguments
refer to relative cohomology over $\IC$, e.g. $H^3=H^3(Z^*,\cxH;\IC)$.}
$H^3$ depends only on the topological 
data, the fiber is up to monodromy constant over $\cx M$, and there 
is a trivially flat connection, $\nabla$, called the Gauss-Manin
connection. The Hodge decomposition $H^3=\oplus_{p=0}^3 H^{3-p,p}$
varies over $\cx M$, as the definition of the Hodge degree
depends on the complex structure. The Hodge filtrations $F^p$ 
$$
H^3=F^0\supset F^1\supset F^2\supset F^3\supset F^{4}=0\ ,\qquad 
F^p=\oplus_{q\geq p}H^{q,3-q}\subset H^3\  ,
$$
define holomorphic subbundles $\cx F^p$ whose fibers are the subspaces
$F^p\subset H^3$. The action of the Gauss-Manin 
connection $\nabla$ on these subbundles has the property 
$
\nabla (\cx F^p) \subset \cx F^{p-1}\otimes T^*_\cx M,
$
known as Griffiths transversality. 

Concretely, the mixed Hodge structure on the relative cohomology space $H^3(Z^*,\cxH)$ 
looks as follows. The Hodge filtrations are
\eqn\hfi{\vbox{\offinterlineskip\halign{
\strut ~$#$~\hfil&~$#$~\hfil &~$#$\hfil \cr
F^3=&H^{3,0}(Z^*,\cxH)&=H^{3,0}(Z^*)\ ,\cr
F^2=&F^3\oplus H^{2,1}(Z^*,\cxH)&=F^3 \oplus H^{2,1}(Z^*)\oplus H_{var}^{2,0}(\cxH)\ ,\cr
F^1=&F^2\oplus H^{1,2}(Z^*,\cxH)&=F^2 \oplus H^{1,2}(Z^*)\oplus H_{var}^{1,1}(\cxH)\ ,\cr
F^0=&F^1\oplus H^{0,3}(Z^*,\cxH)&=F^1 \oplus H^{0,3}(Z^*)\oplus H_{var}^{0,2}(\cxH)\ ,\cr
}}}
where the equations to the right display the split
$H^3(Z^*,\cxH)\simeq{\rm ker\ } \big( H^3(Z^*)\to H^3(\cxH)\big) 
\oplus {\rm coker\ } \big( H^2(Z^*)\to H^2(\cxH)\big)$.
The weight filtration is defined as 
$$
W_2 = 0\ ,\qquad W_3 = H^3(Z^*)\ ,\qquad W_4=H^3(Z^*,\cxH)\ ,
$$
such that the quotient spaces $W_3/W_2 \simeq H^3(Z^*)$ and 
$W_4/W_3 \simeq H^2(\cxH)$ define pure Hodge structures. 
\def\dcl{\de_z}\def\dop{\de_\zh}
Variations in the closed ($\dcl$) and open ($\dop$) string sector
act schematically as  
\eqn\ocdia{
\xymatrix{
F^3 \cap W_3\ar[r]^\dcl\ar[dr]^\dop 
&F^2 \cap W_3 \ar[r]^\dcl\ar[dr]^\dop 
&F^1 \cap W_3 \ar[r]^\dcl\ar[dr]^\dop  
&F^0 \cap W_3  
\cr
&F^2 \cap (W_4/W_3) \ar[r]^{\dcl,\dop}
&F^1 \cap (W_4/W_3)\ar[r]^{\dcl,\dop} 
&F^0 \cap (W_4/W_3)\cr 
}
}

The variation of the Hodge structure over $\cx M$ can be measured by the period matrix
\def\uPi#1#2{\ux\Pi_#1^#2}\def\uXi#1#2{\al_{#2}^{(#1)}}
$$
\uPi A {\Si}=\int_{\ga_\Si} \al_{A}\, , \qquad \al_{A}\in H^3 \ ,
$$
where $\ga_\Si$ is a fixed topological basis for $H_3(Z^*,\cxH)$ and 
$\{\al_{A}\}$ with $A=1,...,\dim(H^3)$ denotes a basis of relative 3-forms. 
One may choose an ordered basis $\{\uXi q {A}\}$ adapted to the Hodge filtration, 
such that the subsets $\{\uXi {q'} {A}\}$, $q'\leq q$ span the spaces $F^{3-q}$ 
for $q=0,..., 3$.

To make contact between the Hodge variation and the $B$-model defined at
a point $m\in \cx M$, the Gauss-Manin connection has to be put into
a form compatible with the chiral ring properties of a SCFT. Chiral operators of
definite $U(1)$ charge are identified with forms of definite Hodge degree,
which requires a projection onto the quotient spaces $F^p/F^{p+1}$ at
the point $m$.
Moreover, the canonical CFT coordinates $t_a$, centered at $m\in \cx M$,
should flatten the connection $\nabla$ and we require
\eqn\gmflat{
\nabla_a \uXi q A (m) = \p_{t_a}\uXi q A(m)\buildrel ! \over =(C_a(t)\cdot \uXi q A)(m)\ \in 
F^{3-q-1}/F^{3-q}|_m\ .
}
The second equation is an important input as it expresses the 
non-trivial fact, that in the CFT, an infinitesimal deformation 
in the direction $t_a$ is generated by an insertion of (the descendant)
of a chiral operator $\phi_a\SS1$ in the path integral, which in turn
can be described by a naive multiplication by the operator $\phi_a\SS1$
represented by the connection matrix $C_a(t)$. 
The above condition assumes that such a simple relation
holds on the 
full open-closed deformation space for all deformations in  $F^2/F^3$. 
Thus  $\phi_a\SS1$ 
can be either a bulk field of left-right $U(1)$ charge $(1,1)$ or a boundary operator of total $U(1)$ charge 1. The consistency of the results obtained below with this ansatz and the correct matching with the CFT deformation space discussed in sect.~6 provides evidence in favor of a proper CFT realization of this structure.

Phrased differently, we consider 
the $\uXi q A $ as flat sections of an ``improved'' flat connection $D_a$ 
in the sense of refs.~\refs{\TTstar,\BCOV}
$$
D_a\uXi q A =0,\qquad D_a=\p_{z_a}-\Ga_a(z)-C_a(z),\qquad [D_a,D_b]\,=\,0 \ ,
$$
where $z_a$ are local coordinates on $\cx M$, 
the connection terms $\Ga_a(z)$ and $C_a(z)$ 
are maps from $\cx F^{3-q}$ to $\cx F^{3-q}$ and 
$\cx F^{3-q-1}$, respectively, and $\Ga_a(z)$ vanishes in the canonical 
CFT coordinates $t_a$.\foot{See sect.~2.6 of ref.~\BCOV\ for the definition 
of canonical coordinates from the (closed-string) CFT point of view.}

\def\uPi#1#2{\ux\Pi_#1^#2}\def\uXi#1#2{\al_{#2}^{(#1)}}\def\Omu{\ux \Om}
Instead of working in generality, we study the Gauss-Manin connection
for the relative cohomology group on the two parameter 
family of branes on the quintic
defined by \defhone. 
We consider a large volume phase with moduli \defmod\ 
defined by the following linear combination of charge vectors \elti:
$$
\tx l^1=(-4;0,1,1,1,1,-1,1),\qquad \tx l^2=(-1;1,0,0,0,0,1,-1)\ ,
$$
A complete set of differential operators derived from eq.~\gkz\ is given by
\eqn\qdiffops{\eqalign{
\cx L_1 &= \th_1^4+(4 z_1 (\th_1-\th_2)-5z_1 z_2(4 \th_1+\th_2+4))\prod_{i=1}^3
(4 \th_1+\th_2+i)\ , \cr
\cx L_2 &=  \th_2(\th_1-\th_2)+z_2 (\th_1-\th_2)(4 \th_1+\th_2+1)\ ,\cr
\cx L_3 &=  
\th_1^3 (\th_1-\th_2)+(4 z_1 \prod_{i=1}^3(4 \th_1+\th_2+i)+z_2\th_1^3)
(\th_1-\th_2)\ .
}}
Computing the ideal generated by the $\cx L_k$ acting on $\Omu$ shows that
$H^3$ is a seven-dimensional space spanned by the
multiderivatives $(1,\th_1,\th_2,\th_1^2,\th_1\th_2,\th_1^3,\th_1^2\th_2)$
of $\Omu$. The dimensions $d_q=\dim(F^{3-q}/F^{3-q+1})$ are 
$1,2,2,2$ for $q=0,1,2,3$, respectively. 
The $d_1=2$ directions tangent to $\cx M$ 
represent the single complex structure deformation $z=z_1z_2$ of the 
mirror quintic $Z^*$ and the parameter $\zh=z_2$ parametrizing
the family of hypersurfaces. 

To implement a CFT like structure at a point $m\in \cx M$, one may take
linear combinations of the multiderivatives acting on $\Omu$ to obtain  
ordered bases $\{\uXi q {A}\}$ and 
$\{\ga_\Si\}$ which bring the period matrix into a block upper triangular 
form\foot{It is understood, that all entries in the following matrices are
block matrices operating on the respective subspaces of definite $U(1)$ charge, 
with dimensions determined by the numbers $d_q$.} 
\def\one{1\hskip-2pt 1}
\eqn\trigperiods{\uPi \Si {A}\,=\,
\pmatrix{ 1 & * & * & * \cr 0 & \one_{d_1\times d_1} & * & * \cr 0 & 0 & \one_{d_2\times d_2} & * \cr 0 & 0 & 0 & \one_{d_3\times d_3} } \ .}
Griffiths transversality then implies that 
in the local coordinates at a point of maximal unipotent monodromy
\eqn\GMCon{
\pmatrix{
\nabla \uXi 0 1 \cr
\nabla\uXi 1 2 \cr
\nabla \uXi 1 3 \cr
\nabla \uXi 2 4 \cr
\nabla  \uXi 2 5 \cr
\nabla \uXi 3 6 \cr
\nabla  \uXi 3 7 }
\,=\,  \pmatrix{
0 & \sum_{a=1}^2\fc{dz_a}{z_a}M_{a}^{(1)} & 0 & 0 \cr
0 & 0 & \sum_{a=1}^2\fc{dz_a}{z_a}M_{a}^{(2)} & 0 \cr
0 & 0 & 0 & \sum_{a=1}^2\fc{dz_a}{z_a}M_{a}^{(3)} \cr
0 & 0 & 0 & 0}
\pmatrix{
\uXi 0 1 \cr
\uXi 1 2 \cr
\uXi 1 3 \cr
\uXi 2 4 \cr
\uXi 2 5 \cr
\uXi 3 6 \cr
\uXi 3 7} \ , }
where the moduli-dependent matrices $M_{a}^{(q)}$ of dimension 
$d_{q-1}\times d_{q}$ are derivatives of the entries of 
$\ux \Pi$ in eq.\trigperiods. The above expression is written in 
logarithmic variables $\log(z_a)$, anticipating the 
logarithmic behavior of the periods at the point of maximal unipotent 
monodromy centered at $z_a=0$. In 
local coordinates $x_a$ centered at a generic point
$m\in \cx M$, the periods are analytic in 
$x_a$ and $dz_a/z_a$ should be replaced with $dx_a$.

The left upper block can be brought into the form
$$
\sum_{a=1}^2\fc{dz_a}{z_a}M_{a}^{(1)}=(\fc{dq_1}{2\pi i\,q_1},\fc{dq_2}{2\pi i\,q_2})
$$
by the variable transformation
\eqn\mima{
q_a(z)=\exp(2\pi i \,\uPi 1 {{a+1}} (z))\ .
}
It has been proposed in ref.~\LMW, that eq.\mima\ represents the mirror map
between the $A$-model K\"ahler coordinates $t_a=\fc{1}{2\pi i}\ln(q_a)$ on the open-closed 
deformation space of an $A$-type compactification $(Z,L)$
and the coordinates $z_a$ on the complex structure moduli space 
of an $B$-type compactification $(Z^*,E)$ near a large complex structure point. 
We propose that the above flatness conditions defines more generally the mirror map between the 
open-closed deformation spaces for any point $m \in \cx M$.\foot{A non-trivial example will be 
described in ref.~\wip.} It is worth stressing that the
mirror map defined by the above flatness argument coincides with the mirror map obtained
earlier in refs.~\refs{\AV,\AKV} for non-compact examples by a physical argument, using 
domain wall tensions and the Ooguri-Vafa expansion at a large complex structure point. 
This coincidence can be viewed as experimental evidence for the existence of a more fundamental 
explanation of the observed flat structure from the underlying topological string theory,
as advocated for in this note.

Identifying $\uXi 0 1$ with the unique operator $\phi\SS0=1$ 
and the $\uXi 1 {a+1}$  with the charge one operators $\phi_a^{(1)}$
associated with the flows parametrized by $\log(q_a)$,
eq.~\GMCon\ implements the CFT relation
$$
\nabla_{q_a} \phi^{(0)} = \phi_a^{(1)} = \phi_a^{(1)}\cdot \phi\SS0 \ ,
$$
discussed below \gmflat.
The above series of arguments and manipulations 
is standard material in closed string mirror symmetry 
and led to the deep connection between the geometric Hodge variations
of CY three-folds in the $B$-model and $A$-model quantum cohomology on the mirror.%
\foot{See refs.~\refs{\mibo,\CK} for background material and a comprehensive list of references.}
After the variable transformation \mima\ and restricting to the
subspace $H^3(Z^*)$ describing the complex moduli space $\cx M_{CS}$
of the mirror quintic, eq.\GMCon\ becomes
\eqn\GMConcl{
\pmatrix{
\nabla \uXi 0 {cl} \cr
\nabla\uXi 1 {cl}{} \cr
\nabla \uXi 2 {cl} \cr
\nabla \uXi 3 {cl} \cr}
\,=\,  \pmatrix{
0 & \fc{dq}{2\pi i \,q}& 0 & 0 \cr
0 & 0 & C(q)\fc{dq}{2\pi i \,q}& 0 \cr
0 & 0 & 0 & -\fc{dq}{2\pi i \,q}\cr
0 & 0 & 0 & 0 }
\pmatrix{
\uXi 0 {cl} \cr
\uXi 1 {cl} \cr
\uXi 2 {cl} \cr
\uXi 3 {cl}}
\ , }
with $\uXi p {cl}\in H^{3-p,p}(Z^*,\IC)$ and 
$q=q_1q_2=e^{2\pi i t}$. Under mirror symmetry these data get
mapped to the K\"ahler volume $t$ and the so-called 
Yukawa coupling $C(q)=5+\cx O(q)$, which  
describes the classical intersection and the Gromov-Witten invariants
on the quintic. In the CFT, the quantities $C(q)$ represent the 
moduli-dependent structure constants of the ring of chiral primaries 
defined in ref.~\LVW.

The point which we are stressing here is that at least part of these concepts continue to 
make sense for the Hodge variation \GMCon\ on the full relative cohomology 
space $H^3(Z^*,\cxH)$ over the open-closed deformation space
$\cx M$ fibered over $\cx M_{CS}$. More importantly, the Hodge theoretic
definition of mirror symmetry described above gives correct results for
the open string analogues of the Gromov-Witten invariants
in those cases, where results have been obtained by different methods, such
as space-time arguments involving domain walls \refs{\AV,\AKV}.%
\foot{See \refs{\LM,\LMW,\Forbes,\JS,\AHMM,\JSii} for various examples.}

In this sense, the existence of a flat structure observed above, 
and the agreement of the Hodge theoretic results with other methods, 
if available, urges for a proper CFT description of 
the deformation families defined over $\cx M$ and an appropriate open-string extension of 
$A$-model quantum cohomology. 
In the following we collect further evidence in favor of 
an interesting integrable structure on the open-closed deformation space, 
working in the $B$-model.

\subsec{Integrability conditions}
The correlation functions of the topological family of closed-string CFTs 
satisfy the famous WDVV integrability condition \refs{\WittenWDVV,\WDVV}. In the context of 
the $B$-model on a CY three-fold, this condition becomes
part of $\cx N=2$ special K\"ahler geometry of the complex structure moduli
space, which implies, amongst others, 
the existence of a single holomorphic prepotential $\cx F$ that
determines all  entries of the period matrix in the canonical CFT 
coordinates $t_a$.

There exists no prepotential 
for the period matrix \trigperiods\ on the relative cohomology group $H^3(Z^*,\cxH)$,
but certain aspects of the $\cx N=2$ special geometry
of the closed-string sector $H^3(Z^*)\subset H^3(Z^*,\cxH)$ generalize to the larger 
cohomology space, justifying the term $\cx N=1$
special geometry \refs{\PM,\LMW}.\foot{$\cx N=1,2$ denotes the
number of 4d space-time supersymmetries of the CY compactification 
of the physical type II string to four dimensions with and without branes.} 
Some aspects of this $\cx N=1$ special geometry have been worked out for
non-compact $Z^*$  in \refs{\LMW,\Forbes} and we add here some missing pieces 
for the compact case. In the following we 
work at a ``large complex structure point'' $m_0\in \cx M$
of maximal unipotent monodromy. 
The existence of such points $m_0$ follows from the 
general property of the GKZ systems described in sect.~3.

We start from the following general ansatz for the 7-dimensional period vector 
of the holomorphic 3-form
\eqn\pvsi{
\uPi 1 \Si = \big(1,\ t,\ \hx t,\ F_t(t),\ W(t,\that),\ F_0(t),\ T(t,\that)\ \big)\ ,
}
where $t$ is the closed- and $\hx t$ the open-string deformation, related
to the flat normal crossing divisor coordinates $(t_1(z_1,z_2),t_2(z_1,z_2))$ 
of eq.~\mima\ by the linear transformation $t=t_1+t_2$, $\hx t=t_2$. The subset of 
periods in the closed-string sector is determined by the prepotential 
$\cx F$ as $(1,t,F_t=\p_t\cx F,F_0(t)=2\cx F(t)-t\p_t \cx F)$ and depends only 
on $t$. The additional periods  $(\that, W(t,\that), T(t,\that))$ are so far 
arbitrary functions, except that the leading behavior at $m_0$ at $z_a=0$ is,
schematically,
$$
t,\that\sim \log(z_.),\qquad F_t,W \sim \log^2(z_.),\qquad F_0,T \sim \log^3(z_.) \ .\qquad 
$$
The function $W$ is in some sense the closest analogue of the closed
string prepotential and indeed has been conjectured to be
a generating function for the open-string disc invariants in \refs{\LMW,\JS}.

For an appropriate choice of basis $\{\uXi q A\}$, the 
period matrix takes the upper triangular form \trigperiods\ with entries
\eqn\pvsii{
(\ux \Pi)=\pmatrix{1&t&\that&F_t&W&F_0&T\cr
0&1&0&F_{t,t}&W_{,t}&F_{0,t}&T_{,t}\cr 
0&0&1&0&W_{,\that}&0&T_{,\that}\cr
0&0&0&1&0&-t&\mu\cr
0&0&0&0&1&0&\rho\cr
0&0&0&0&0&1&0\cr
0&0&0&0&0&0&1\cr
}\ , 
}
where the derivatives w.r.t. $t$  and $\that$ are denoted by subscripts and
the functions $\mu$ and $\rho$ are defined by 
$$
\mu=\fc{W_{,t\that}T_{,tt}-W_{,tt}T_{,t\that}}{CW_{,t\that}},\qquad
\rho=\fc{T_{,t\that}}{W_{,t\that}},\qquad
C=\cx F_{,ttt}=F_{t,tt}\ .
$$
The connection matrices read
\eqn\GMmat{\eqalign{
M_t\,=\, \pmatrix{
0 & 1 & 0 & 0 & 0 & 0 & 0 \cr
0 & 0 & 0 & C & W_{,tt} & 0 & 0 \cr
0 & 0 & 0 & 0 & W_{,t\that} & 0 & 0 \cr
0 & 0 & 0 & 0 & 0 & -1 & \mu_{t} \cr
0 & 0 & 0 & 0 & 0 & 0 & \rho_{,t} \cr
0 & 0 & 0 & 0 & 0 & 0 & 0 \cr
0 & 0 & 0 & 0 & 0 & 0 & 0},\ 
M_\that\,=\,\pmatrix{
0 & 0 & 1 & 0 & 0 & 0 & 0 \cr
0 & 0 & 0 & 0 & W_{,t\that} & 0 & 0 \cr
0 & 0 & 0 & 0 & W_{,\that\that} & 0 & 0 \cr
0 & 0 & 0 & 0 & 0 & 0 & \mu_{,\that} \cr
0 & 0 & 0 & 0 & 0 & 0 & \rho_{,\that} \cr
0 & 0 & 0 & 0 & 0 & 0 & 0 \cr
0 & 0 & 0 & 0 & 0 & 0 & 0 } \ . }}
The integrability condition $\p_tM_\that-\p_\that M_t+[M_\that,M_t]=0$
implies that 
\eqn\Treln{
\rho(t,\that)=aW_{,\that}+b\ ,\qquad \mu(t,\that)=aC^{-1}(t)\left(\int  
( W_{,t\that}^2-W_{,tt}W_{,\that\that})\, d\that +g(t)\right) \ , 
}
with $a,b$ some complex constants and $g(t)$ an undetermined function. 
The relation \gmflat\ then implies 
that the period $T$ is of the form
\eqn\Trel{
T(t,\that)=\int \left(\fc{a}{2}W_{,\that}^2+bW_{,\that}\right) d\that +f(t)\ ,
}
with $\p_t^2f(t)=g(t)$. The integrability condition \Trel\ determines
the top period in the open-string sector in terms of the other periods,
up to the function $f(t)$. In this sense it is similar to the relation
in the closed-string sector, that determines the top period $F_0$ in terms of the 
other periods. The integration constants can be fixed by determining
the leading behavior of the periods in the large volume limit, as we will
do in app.~A for the quintic example.

The above argument and the integrability relation \Trel\ 
applies to any two parameter family with one closed- and one 
open-string modulus and can be straightforwardly generalized to more parameter
cases. For a given geometry, such as the quintic family 
described by the operators $\cx L_k$ in \qdiffops, one can of course  reach the same conclusion
by studying the explicit solutions and also 
determine the function $f(t)$.  As noticed
below eq.~\bdrycont, the relative forms in the open-string sector of this family 
can be associated with the Hodge variation on a quartic K3 surface. The period
vector $\vx\pi$ of the K3 surface is spanned by the solutions 
$\p_\that(\that,W,T)=(1,W_{,\that},T_{,\that})$ and
the integrability condition \Trel\ represents an algebraic relation
$\vx \pi^T\, \hx \eta\, \vx \pi=0$ amongst the K3 periods, where $\hx \eta$ 
is the intersection matrix. 
We come back to the intersection form $\hx \eta$ when we discuss the 
topological metric in sect.~7\yyy.

The above discussion was essentially independent of the choice of the
large complex structure point $m_0$ and a similar argument for other $m$ shows 
that the integrability condition $\Trel$
holds for any $m\in \cx M$ in the local coordinates defined by \mima.

\subsubsec{A cautious remark}
The similarities of the above arguments with those in the case 
of closed string mirror symmetry may have obscured the fact that one crucial datum 
is still incomplete: the topological metric $\eta$ on the open-closed state space. 
In the closed-string sector,
the topological metric $\eta_{cl}$ 
is given by the classical intersection matrix on $H_3(Z^*)$ and  
its knowledge permits, amongst others, the 
determination of the true geometric periods as a particular linear combinations 
of the solutions to the Picard-Fuchs equations.
More importantly, the topological metric is needed 
to complete the argument that identifies $C(q)$ in eq.\GMConcl\ with
the structure constants of the chiral ring of ref.~\LVW, as well as to 
access the non-holomorphic and higher genus sector of the theory 
using the $tt^*$ equations 
\TTstar\ and the holomorphic anomaly equation \BCOV. 
Our present lack of a precise understanding of the topological metric in the open-string
sector renders the following sections somewhat fragmentary.

In the next section we derive predictions for disc invariants by fixing 
the geometric periods in a way that avoids the knowledge of the topological 
metric.
Some observations on the other issues mentioned above, including a proposal 
for the full topological metric, will be discussed in sects.~6 and 7.

\newsec{Large radius invariants for the $A$-model}
The geometry of $A$-branes is notorically more difficult to study 
than that of the $B$-branes. For the type of $B$-branes studied above,
the $A$-model mirror geometry can be in principle 
obtained from the toric framework of ref.~\AV.
The conjectural family of $A$-branes mirror to the $B$-brane family 
studied above, is defined on the quintic hypersurface $Z$ in the toric ambient space 
$W=\cx O(-5)_{\IP^4}$ with homogeneous coordinates
$$
\matrix{x_0&x_1&x_2&x_3&x_4&x_5\cr -5&1&1&1&1&1}
$$
The K\"ahler moduli ($t_1,t_2)$ mirror to the complex structure 
coordinates $(z_1,z_2)$ are defined by the equations
\eqn\defla{\eqalign{
&\sum_{i=1}^5|x_i|^2-5|x_0|^2=\Im t=\Im t_1+\Im t_2,\qquad |x_1|^2-|x_0|^2=\Im t_2\ ,\cr
}}
with $\Im t_1,\Im t_2>0$. The first constraint holds on all of $Z$ and the
K\"ahler modulus $t$ describes the closed string deformation, the overall
K\"ahler volume of $Z$. The second constraints holds only on the 
Lagrangian submanifold $L$ and describes the open-string deformation.

The toric framework of ref.~\AV\ gives an explicit description of the geometry 
of Lagrangian subspaces in the ambient space $W$, which has been used to 
study an interesting class of non-compact branes for CY ambient spaces, see e.g.
\refs{\AV,\AKV,\LMW}.
The clear geometric picture of the toric description is lost for hypersurfaces
and the searched for subspace
$L\subset Z$ carrying the mirror $A$-brane has no simple description, 
at least at general points in the (full) moduli space and to our knowledge.
However, by the homological mirror symmetry conjecture \Kont, we expect that
a corresponding $A$-brane, which is mirror to the $B$-brane given in terms of the discussed
divisor together with its curvature 2-form, should be present in the $A$-model
geometry. Clearly, in order to complete our picture a constructive recipe of mapping $B$-branes
to the corresponding $A$-branes for compact CY geometries is desirable. We hope to come back to this
issue elsewhere.

Since the $A$-model geometry is naively independent of the complex structure
moduli, one is tempted to choose a very special form of the hypersurface 
constraint to simplify the geometry. 
In ref.~\katzi\ it is shown how the number 2875 of lines in the generic 
quintic can be determined from the number of lines in a highly degenerate
quintic,
defined by the hypersurface constraint  
\eqn\dfi{
P(Z_\al) = p_1.p_4+\al\, p_5,\qquad  \al\in \Delta\ .
}
Here $\al$ is a parameter on the complex disc $\Delta$ and the $p_k$ are degree $k$
polynomials in the homogeneous coordinates of $\IP^4$. At the point 
$\al=0$ the quintic splits into two components of degree one and four. 
Katz shows, that there are 1600+1275=2875 holomorphic maps to 
lines in the two components of the central fiber that deform to 
the fiber at $\al\neq 0$. 

The $N_1=2875$ curves of degree one 
contribute to the tension $\cx T$ of a D4-brane wrapping
the 4-cycle $\Ga = H\cap Z$ as
$$
\cx T=-\fc{5}{2}t^2+\fc{1}{4\pi^2}\left(2875 \sum_k \fc{q^{k}}{k^2}+2\cdot609250 \sum_k \fc{q^{2k}}{k^2}+... \right)
$$
where $q=\exp(2\pi i t)$, $H$ is the hyperplane class and 
the dots denote linear and constant terms in $t$ as well 
as instanton corrections
from maps of degree $d>2$.  
In the singular CY the generic 
4-cycle splits into two components and one expects two 
separate contributions 
$$
\cx T\SS1= c\SS1\, t^2+\fc{N\SS1_d}{4\pi^2} \sum_k \fc{q^{dk}}{k^2}+...,\qquad
\cx T\SS2= \cx T-\cx T_1\ ,
$$
with $N\SS1_1=1600$ and $N\SS2_1=1275$.

As explained in ref.~\katzi, there are other genus zero maps 
to the two components, that develop nodes 
at the intersection locus $p_1=p_4=0$ upon deformation, and they 
do not continue to exist as maps from $S^2$ to $S^2$. The idea is 
that in the presence of the Lagrangian $A$-brane on the degenerate
quintic, the nodes of the spheres can open up to become 
the boundary of holomorphic disc instantons ending on $L$.
Indeed the two independent
double logarithmic solutions of the Picard-Fuchs system \qdiffops\ can be 
written in the flat coordinates \mima\ as
\eqn\tensions{\eqalign{
\cx T\SS1&= -2 t^2+
\fc{1}{4\pi^2}\sum_k\ \fc{1}{k^2}\big(1600 q^{k}+2\cdot 339800 q^{2k}+...\big)
+\cx T\SS o(t_1,t_2) \cr
\cx T\SS2&= -\fc{1}{2} t^2+\fc{1}{4\pi^2}
\sum_k\ \fc{1}{k^2} \big( 1275q^{k}+2\cdot 269450 q^{2k}+...\big)
-\cx T\SS o(t_1,t_2) 
}
}
showing the expected behavior and adding up to the
closed-string period $\cx T$. The split of the degree two curves, 
$N_2=609250=339800+269450=(258200+\h 163200)+(187850+\h 163200)$
is compatible with the results of ref.~\katzii.

The extra contribution $\cx T\SS o(t_1,t_2)$ can be written as
$$
\cx T\SS o(t_1,t_2)= 4tt_2-2t_2^2+\fc{1}{4\pi^2}\sum_{k,n_1,n_2 \atop n_1\neq n_2}\ \fc{1}{k^2}\ N_{n_1,n_2}(q_1^{n_1}q_2^{n_2})^k\ .
$$
The first few coefficients $N_{n_1,n_2}$ for small 
$n_i$, including the contributions from $n_1=n_2$, are listed in table~1 below.

\vbox{
$$\hskip-1cm\vbox{\offinterlineskip\halign{
\hfil~$#$~&\hfil~$#$~&\hfil~$#$~
&\hfil~$#$~&\hfil~$#$~&\hfil~$#$~
&\hfil~$#$~&\hfil~$#$~&\hfil~$#$~
&\hfil~$#$~&\hfil~$#$~&\hfil~$#$~
&\hfil~$#$~&\hfil~$#$~&\hfil~$#$~
&\hfil~$#$~\vrule\cr
_{n_2}\setminus ^{n_1}\vr & 0&1&2&3&4&5\cr
\noalign{\hrule}
0\ \vr&0& -320& 13280& -1088960& 119783040& -15440622400\cr 1\ \vr&20& 1600& -116560& 
12805120& -1766329640& 274446919680\cr 2\ \vr&0& 2040& 679600& -85115360& 
13829775520& -2525156504560\cr 3\ \vr&0& -1460& 1064180& 530848000& -83363259240&
16655092486480\cr 4\ \vr&0& 520& -1497840& 887761280& 
541074408000& -95968626498800\cr 5\ \vr&0& -80& 1561100& -1582620980& 
931836819440& 639660032468000\cr 6\ \vr&0& 0& -1152600& 
2396807000& -1864913831600& 1118938442641400\cr 7\ \vr&0& 0& 580500& -2923203580&
3412016521660& -2393966418927980\cr 8\ \vr&0& 0& -190760& 
2799233200& -5381605498560& 4899971282565360\cr 9\ \vr&0& 0& 37180& -2078012020& 
7127102031000& -9026682030832180\cr 10\ \vr&0& 0& -3280& 
1179935280& -7837064629760& 14557931269209000\cr 11\ \vr&0& 0& 0& -502743680& 
7104809591780& -20307910970428360\cr 12\ \vr&0& 0& 0& 155860160& -5277064316000& 
24340277955510560\cr 13\ \vr&0& 0& 0& -33298600& 
3187587322380& -24957649473175420\cr 14\ \vr&0& 0& 0& 4400680& -1549998228000& 
21814546476229120\cr 15\ \vr&0& 0& 0& -272240& 
597782974040& -16191876966658500\cr 16\ \vr&0& 0& 0& 0& -178806134240& 
10157784412551120\cr 17\ \vr&0& 0& 0& 0& 40049955420& -5351974901676280\cr
\noalign{\hrule}
}}
$$
\nobreak\tablecaption{1}{\hskip1cm Predictions for Ooguri--Vafa invariants.}}
\vskip0.5cm

According to the general philosophy of the Hodge theoretic mirror map 
described in the previous sections, the double logarithmic solutions 
represent the generating function of holomorphic discs ending on the 
$A$-brane $L$. In the basis of sect.~4\yyy\ we find 
$$
F_t=\cx T\SS1+\cx T\SS2=\cx T, \qquad W=\cx T\SS1\ .
$$
Assuming that the normalization argument leading to \tensions\ is correct, 
the numbers $N_{n_1,n_2}$ of table 1 
are genuine Ooguri-Vafa invariants for the $A$-brane geometry predicted
by mirror symmetry.
It would be interesting to justify the above arguments and the
prediction for the disc invariants in table 1
by an independent computation. Further evidence for the above results 
will be given in sect.~7 and app.~A, by deriving the same result from the 
afore mentioned duality to Calabi--Yau four-folds.

\newsec{Relation to CFT correlators} 
The relevant closed-string observables in the BRST cohomology of the topological B-model of a Calabi-Yau manifold are locally given by \WittenZZ
\eqn\closedob{
\phi^{(p)}\,=\,\phi^{(p)}{}^{j_1\cdots j_p}_{\bar\imath_1\cdots \bar\imath_p}
\eta^{\bar\imath_1}\cdots\eta^{\bar\imath_p}\theta_{j_1}\cdots\theta_{j_p} \ , }
where the world-sheet fermions, $\eta^{\bar\imath}=\psi^{\bar\imath}_++\psi^{\bar\imath}_-$ and $\theta_i=g_{i\bar\jmath}\left(\psi^{\bar\jmath}_+-\psi^{\bar\jmath}_-\right)$, are sections of the pullbacks of the anti-holomorphic tangent bundle and the holomorphic cotangent bundle of target-space Calabi-Yau manifold. For the Calabi-Yau three-fold $Z^*$ these observables $\phi^{(p)}$ are identified geometrically with representatives in the sheaf cohomology groups 
\eqn\closedobgeom{ \phi^{(p)} \in H^p(Z^*,{\Lambda}^pTZ^*)\simeq H^{(3-p,p)}(Z^*) \ , \qquad p=0,1,2,3 \ . }  
The last identification is due to the contraction with the unique holomorphic (3,0) form of the Calabi-Yau three-fold $Z^*$. The integer $p$ represents the left and right $U(1)$ charge of the bulk observable $\phi^{(p)}$. 

The local open-string observables for a worldsheet with B-type boundary are analogously given by
\eqn\openob{
\hat\phi^{(p+q)}\,=\,{\hat\phi^{(p+q)}}{}^{j_1\cdots j_q}_{\bar\imath_1\cdots \bar\imath_p}
\eta^{\bar\imath_1}\cdots\eta^{\bar\imath_p}\theta_{j_1}\cdots\theta_{j_q} \ . }
In the absence of a background gauge field on the worldvolume of the brane the fermionic modes $\theta_j$ vanish along Neumann directions whereas the fermionic modes $\eta^{\bar\imath}$ vanish along Dirichlet directions on the boundary of the worldsheet \HoriCK. Hence, locally we view the fermionic modes $\theta_j$ as sections of the normal bundle and the fermionic modes $\eta^{\bar\imath}$ as sections of the anti-holomorphic cotangent bundle of the brane. With background fluxes on the brane worldvolume the boundary conditions become twisted and obey \WitCS 
\eqn\bdryflux{ \theta_i = F_{i\bar\jmath}\eta^{\bar\jmath} \ . }
In ref.~\KatzGH\ it is explicitly demonstrated that the observables~\openob\ in the BRST~cohomology of the open-string sector for a brane $E$ arise geometrically as elements of the extension groups
\eqn\ExtObs{ \hat\phi^{(p+q)} \in {\rm Ext}^{p+q}(E,E) \ , \qquad p+q=0,1,2,3 \ .}
In the present context, the integer $p+q$ is equal to the total $U(1)$ charge of the open-string observable $\hat\phi^{(p+q)}$. 

Deformations of the topological B-model are generated by the marginal operators, which correspond to BRST observables with $U(1)$ charge one, and hence they appear in the cohomology groups $H^{(2,1)}(Z^*)$ and ${\rm Ext}^1(E,E)$ for the closed and open deformations, respectively. 

In order to make contact with the Hodge filtration of $H^3(Z^*,\cxH)$ we interpret the divisor $\cxH$ of the Calabi-Yau three-fold $Z^*$ as the internal worldvolume of a B-type brane. For a divisor the extension groups \ExtObs\ simplify \KatzGH, and in particular ${\rm Ext}^1(\cxH,\cxH)$ reduces to $H^0(\cxH, N\cxH) \simeq H^{(2,0)}(\cxH)$, where the last identification results again from the contraction with the holomorphic (3,0) form. For our particular example the cohomology groups $H^{(2,1)}(Z^*)$ and $H^{(2,0)}(\cxH_1)$ are both one-dimensional and therefore are generated by the closed- and open-string marginal operators $\phi$ and $\hat\phi$
$$ \phi^{(1)} \in H^{(2,1)}(Z^*) \subset F^2/F^3 \ , \qquad \hat\phi^{(1)} \in H^{(2,0)}(\cxH_1) \subset F^2/F^3  \ . $$
Due to the identification $F^2/F^3 = H^{(2,1)}(Z^*,\cxH_1) \simeq H^{(2,1)}(Z^*) \oplus H^{(2,0)}(\cxH_1)$ we observe that the infinitesimal deformations $\nabla_t\alpha^{(0)}_1\sim \phi^{(1)}$ and $\nabla_{\hat t}\alpha^{(0)}_1\sim\hat\phi^{(1)}$ in eq.~\gmflat\ precisely agree with the closed and open marginal operators $\phi^{(1)}$ and $\hat\phi^{(1)}$. As a consequence the discussed Picard-Fuchs equations, governing the Hodge filtration $F^p$, describe indeed the deformation space associated to the closed and open marginal operators $\phi^{(1)}$ and $\hat\phi^{(1)}$. 

In the presence of B-type boundaries infinitesimal deformations are generically obstructed at higher order. These obstructions are encoded in the moduli-dependent superpotential generated by disc correlators with insertions of bulk and boundary marginal operators \refs{\LazaroiuNM,\DouglasFR,\GovindarajanUY,\JockersNG}. The relevant disc correlators arise from non-trivial ring relations involving marginal operators and the (unique) boundary top element $\hat\phi^{(3)}\in {\rm Ext}^3(\cxH,\cxH)$. Hence the superpotential is extracted by identifying the element $\hat\phi^{(3)}$ in the relative cohomology group $H^3(Z^*,\cxH)$. For the family of hypersurfaces $\cxH$ the extension group ${\rm Ext}^3(\cxH,\cxH)$ becomes \KatzGH
$$ \hat\phi^{(3)} \in {\rm Ext}^3(\cxH,\cxH) \simeq H^2(\cxH,N\cxH) \simeq H^{(2,2)}(\cxH) \ , $$
where locally $\hat\phi^{(3)}=\hat\phi^{(3)}{}_{\bar\imath\bar\jmath}^{k}\eta^{\bar\imath}\eta^{\bar\jmath}\theta_k$. It is obvious that the cohomology group $H^{(2,2)}(\cxH)$ does not appear in the filtration $F^p$ of the relative cohomology group $H^3(Z^*,\cxH)$. On the other hand the variation of mixed Hodge structure encodes by construction the ring relations of the observables generated by the marginal operators $\phi^{(1)}$ and $\hat\phi^{(1)}$. Therefore we conclude that these marginal operators do not generate the boundary-boundary top element $\hat\phi^{(3)}$. Thus the analyzed deformation problem is unobstructed and does not give rise to a non-vanishing superpotential.

{}From a physics point of view the family of divisors $\cxH$ describes a family of holomorphic hypersurfaces, which all give rise to supersymmetric B-brane configurations, and hence we should not expect any obstructions resulting in a superpotential. 

However, the result of the above analysis drastically changes as we add a $D5$-brane charge on 2-cycles in $\cxH$, e.g. by adding non-trivial background fluxes on the worldvolume of the B-type brane. From a space-time perspective \refs{\LustBD,\JockersZY}, we expect the appearance of F-terms precisely for those two-form background fluxes, whose field strength takes values in the variable cohomology of the hypersurface $\cxH$ 
\eqn\Fflux{ F \in  {\rm coker\ } \big( H^2(Z^*,\IZ)\to H^2(\cxH,\IZ)\big) \ . }
These fluxes induce a macroscopic superpotential 
\refs{\LustBD,\JockersZY,\GMM,\Ma}
\eqn\Ffluxii{
W=\int_D F \wedge \om = \int_\Ga F\wedge \Om,
}
where $\om\in H^{2,0}(D)$ is obtained 
by contracting the bulk (3,0) form $\Om$ with a section of the normal bundle to $D$. 
The second expression, derived in a more general context in ref.~\Ma, 
is equivalent to the first one for an appropriate
choice of 5-chain with boundary $D$.

In the microscopic worldsheet description the worldvolume flux $F_{i\bar\jmath}$ yields twisted boundary conditions~\bdryflux, and the fermionic modes $\theta_i$ of the open-string observables~\openob\ are in general no longer sections of the normal bundle $N\cxH$. Instead they should be viewed as appropriate section in the restricted tangent bundle, $TZ^*|_{\cxH}$ \KatzGH. As a consequence we can trade (without changing the $U(1)$~charge) fermionic modes $\eta^{\bar\jmath}$ with appropriate fermionic modes $\theta_i$. As a result the boundary top element $\hat\phi^{(3)}$ can now be associated with an element in the variable two-form cohomology
\eqn\GenExtTwo{
{\rm Ext^3}(\cxH,\cxH) \ni
{\hat\phi}_{\bar\imath\bar\jmath}^{(3)k} \eta^{\bar\imath}\eta^{\bar\jmath}\theta_k \,{\buildrel F_{i\bar\jmath}\over\longleftrightarrow} \, 
{\hat\phi}^{(3)jk}_{\bar\imath}  \eta^{\bar\imath}\theta_j\theta_k  \, {\buildrel \Omega_{ijk}\over\longleftrightarrow}\, 
{\hat\phi}_{i\bar\jmath}^{(3)}\,dx^i\wedge dx^{\bar\jmath} \in {\rm coker\ } \big( H^2(Z^*)\to H^2(\cxH)\big)  \ . }
Thus in the presence of worldvolume background fluxes the boundary top element $\hat\phi^{(3)}$ {\it does} correspond to an element in the Hodge structure filtration $F^p$, and the superpotential is described by a solution of the Picard-Fuchs equations. In this way the a priori unobstructed deformation problem of divisors $\cxH$ in the Calabi-Yau three-fold is capable to describe superpotentials associated to $D5$-brane charges in $H_2(\cxH)$ \refs{\LMW,\JS,\AHMM}. 

On the other hand, since the discussed F-term fluxes~\Fflux\ are elements of the variable cohomology of the hypersurface $\cxH$, {\it i.e.} the field strength of the fluxes can be extended to exact two forms in the ambient Calabi-Yau space, they do not modify the $D5$-brane K-theory charges. Therefore if a suitable $D5$-brane interpretation is applicable the flux-induced superpotentials describe domain-wall tensions between pairs of $D5$-branes, which wrap homologicaly  equivalent two cycles.

When written in the flat coordinates $t_a=\fc{1}{2\pi i}\ln(q_a)$ in \mima,
the Gauss-Manin connection on the total cohomology space 
takes the form:
\eqn\GMConii{
\pmatrix{
\nabla \uXi 0 1 \cr
\nabla\uXi 1 2 \cr
\nabla \uXi 1 3 \cr
\nabla \uXi 2 4 \cr
\nabla \uXi 2 5 \cr
\nabla \uXi 3 6 \cr
\nabla  \uXi 3 7 }
\,=\,  \sum_b \pmatrix{
0 & C\SS0 _b(q_a)\fc{dq_b}{q_b} & 0 & 0 \cr
0 & 0 & C\SS1 _b(q_a)\fc{dq_b}{q_b}& 0 \cr
0 & 0 & 0 & C\SS2 _b(q_a)\fc{dq_b}{q_b} \cr
0 & 0 & 0 & 0 }
\pmatrix{
\uXi 0 1 \cr
\uXi 1 2 \cr
\uXi 1 3 \cr
\uXi 2 4 \cr
\uXi 2 5 \cr
\uXi 3 6 \cr
\uXi 3 7} \ . }
The most notable difference to the closed-string case (cf. eq.\GMConcl)
is that, whereas the matrix $C_b\SS 0$ still 
is of the canonical form $(C_b\SS 0)_1^{\ l}=\delta_{bl}$,
the matrices $C_b\SS q$ are now both moduli dependent for $q=1,2$:
$$
(C_t\SS 1)=\pmatrix{C&W_{,tt}\cr0&W_{,t\that}\cr},\ 
(C_\that\SS 1)=\pmatrix{0&W_{,t\that}\cr0&W_{,\that\that}\cr},\
(C_t\SS 2)=\pmatrix{-1&\mu_{,t}\cr0&\rho_{,t}\cr},\
(C_\that\SS 2)=\pmatrix{0&\mu_{,\that}\cr0&\rho_{,\that}\cr},\
$$
\noi In correspondence with the closed-string sector it is tempting to 
interprete
the $d_{q-1}\times d_q$ matrices $C_b\SS q $ as the structure
constants of a ring of open and closed chiral operators
$$
\phi\SS1_b\cdot \phi\SS q _k \ \buildrel ? \over = \ (C_b\SS q)_{k}^{\ l}\phi\SS{q+1}_l \ , 
$$
as described in \refs{\LMW,\WLlec}. A rigorous CFT derivation of such a relation
is non-trivial, as the 
Hodge variation mixes bulk and boundary operators and describes 
the bulk-boundary ring in the sense of ref.~\KatzGH, 
about which little is known in the context of topological strings (see however
refs.~\refs{\LazaroiuRK,\HerbstJP}). A related complication is the need of a  
topological metric on the space of closed and open BRST states that mixes 
contributions at different order of the string coupling. The most direct 
way to connect the closed-string periods with a CFT quantity is the 
interpretation as overlap functions between boundary states and chiral 
operators \OOY, and it is likely that a similar idea can be applied to the entries 
of the relative period matrix. It would be interesting to make this precise.
It would also be interesting to understand more generally the relation 
of the above concepts to the CFT results obtained from 
matrix factorizations in refs.~\refs{\BBG,\KSi,\KSii,\Kiii}.

\newsec{F-theory four-folds and effective $\cx N=1$ supergravity}
The four-dimensional effective action for the brane compactification 
$(Z,L)$, or its mirror $(Z^*,E)$, should be described by a general 
$\cx N=1$ supergravity theory, which depends on the K\"ahler potential 
and the superpotential through the function \sugra 
\eqn\gfun{
\cx G =K(z,\zh;\bb z,\bb \zh)+\ln W(z,\zh)+\ln \bb W(\bb z,\bb \zh) \, .
}
Here $(z,\zh)$ denote again local complex coordinates on the bundle 
$\cx M \to \cx M_{CS}$. Similarly the construction of ref.~\AHMM\ associates 
to the mirror pair of brane compactifications
$(Z,L)$ and $(Z^*,E)$ a dual
four-fold compactification with the same supersymmetry. Some details
of this duality, including a lift to a full F-theory compactification,  
will be explained below. Subsequently we 
compare the effective couplings in these two descriptions and 
show that they lead to a consistent proposal for the $\cx N=1$ supergravity function 
$\cx G(z,\zh;\bb z,\bb \zh)$. 

\subsec{Four-fold dualities and F-theory}
The construction of refs.~\refs{\PM,\AHMM}\ associates a toric four-fold polyhedron $\D$ to a 
toric brane configuration defined as in ref.~\AV. For the quintic example, $\D$ is 
defined in \defverts\ and has the property that the toric GKZ system associated
with $\D$ reproduces the GKZ system for the relative cohomology group 
for the brane compactification $(Z^*,E)$ via eq.\gkz. In the following
we describe some details of the duality map on the target and moduli spaces.

First note that the polyhedron $\D$ and its dual polyhedron $\Ds$ actually 
define a mirror pair of CY four-folds $X$ and $X^*$. As is clear from the 
derivation of the Picard-Fuchs equations in sect.~3, 
the relative periods of the brane compactification $(Z^*,E)$ are identified
with the periods of holomorphic $(4,0)$ form on the four-fold 
$X^*$. It follows that the complex structure deformations
of the brane compactification $(Z^*,E)$ map to the complex structure of 
the four-fold $X^*$. Adding mirror symmetry, one obtains the following relation 
between the different compactifications:
\eqn\ffdiai{
\xymatrix{
{\buildrel {\displaystyle(Z,L)}\over {\rm (A-branes)}\hskip 20pt }\ar[d]_{4f}^{dual} \ar[r]^{mirror}_{symmetry} 
&
\ar[l]\hskip 20pt{\buildrel {\displaystyle(Z^*,E)}\over {\rm (B-branes)}}\ar[d]_{4f}^{dual} \cr   
\hskip 25pt{\buildrel {\displaystyle X}\over {\rm { }  }} \hskip 20pt\ar[r]^{mirror}_{symmetry}&\hskip 20pt \ar[l]X^*}
}
The vertical maps in this diagram preserve, whereas the horizontal maps exchange,
the notion of complex and K\"ahler moduli. 

The mirror pair $(X,X^*)$ of four-folds constructed in this way has a very special geometric
structure that reflects the mirror symmetry between $A$-type and $B$-type branes on 
the mirror pair $(Z,Z^*)$ of CY three-folds. Whereas the correspondence between 
{\it moduli spaces} is manifest on the $B$-type side 
as the relation between the periods of $(Z^*,E)$ and $X^*$,\foot{An explicit match of period integrals for a class of examples can be found in ref.~\PM.} there is a simple  
correspondence between the {\it target spaces} on the $A$-type side. Namely,
the mirror four-fold $X$ is a fibration over the complex plane 
\eqn\diafib{
\xymatrix{
Z \ar[r] &X \ar[d]^\pi\cr
&\IC\cr}
}
with generic fiber a CY three-fold of type $Z$ and a degenerate
central fiber at the origin specified by the toric polyhedron constructed in
refs.~\refs{\PM,\AHMM}.
Thus the dual four-fold $X^*$ that captures 
the relative periods of the brane compactification
$(Z^*,E)$ is effectively constructed by fibering the CY three-fold $Z$ for the $A$-branes 
over $\IC$
and then taking the (four-fold) mirror of the fibration $X$ obtained in this way.

The mirror pair $(X,X^*)$ of non-compact CY four-folds can be 
related to a honest four-dimensional F-theory 
compactification by a simple $\IP^1$ compactification of the non-compact 
base of $X$. In this way one obtains a mirror pair of compact 
CY four-folds $(X_c,X_c^*)$, where $X_c^*$ is the four-fold for F-theory compactification.
\eqn\ffdiax{
\xymatrix{
Z \ar[r] &\ar[d]^\pi  \hskip 12pt X_c  \hskip 12pt \ar[r]^{\hskip-20pt mirror}_{\hskip-20pt symmetry}  
&
\ar[l]\hskip 10pt  X_c^*\ \ {\rm (F-theory)}\cr   
&\ \IP^1
}
}
An important point is to identify to the image of the 
large base limit of $X_c$ in the moduli 
space of the F-theory compactification on $X_c^*$, which can be 
deduced from the mirror map and the methods of refs.~\refs{\bmff,\BerglundDM}.
The result is that the large 
volume limit $\Im S=Vol(\IP^1)\to\infty$ maps under mirror symmetry to a
weak coupling limit $g_s\to 0$ 
\eqn\wcl{
Vol(\IP^1)\sim 1/g_s\ \to\infty \ .
}
Thus the 
pair of four-folds $(X,X^*)$ is recovered in the 
decompactification/weak-coupling limit 
and the diagram
\ffdiai\ is completed downwards to 
\eqn\ffdiaii{
\xymatrix{
\hskip 20pt X  \hskip 20pt \ar[r]^{mirror}_{symmetry} 
&
\ar[l]\hskip 20pt  X^*\hskip 20pt   \cr   
\ar[u]^{Vol(\IP^1)\to \infty\ }  \hskip 20pt X_c \hskip 20pt \ar[r]^{mirror}_{symmetry}&\hskip 10pt  \ar[l] X_c^*\ar[u]_{\ g_s\to 0}
}
}
On the one hand, the details of the $\IP^1$ compactification determine 
the subleading corrections in $g_s$ but become irrelevant in the 
decompactification/decoupling limit. On the other hand it is worth
stressing, that the ``duality'' between non-compact Calabi--Yau four-folds
$(X,X^*)$ and the mirror pair of brane compactifications 
$(Z,L)$ and $(Z^*,E)$, which underlies the superpotential computation,
represents the {\it strict} limit $g_s=0$, where 
most of the degrees of freedom decouple from the superpotential sector. 
A true duality can exist, if at all, only at the 
level of the lower row of the above diagram.

In the concrete example of sect.~4, the $\IP^1$ compactification can be
obtained by adding the extra vertex 
\eqn\evert{
\tx \nu _8=(1,0,0,0,-1)
}
to the toric polyhedron \defverts\ of the non-compact CY four-fold $X$. This 
defines a compact Calabi--Yau four-fold $X_c$. The K\"ahler modulus $S$ of the 
compact $\IP^1$ base is described by the charge vector 
$$
\tx l^3=(0;-2,0,0,0,0,0,1,1).
$$
The mirror manifold $X_c^*$ is elliptically fibered and defines an F-theory compactification
that will be used to compute the effective couplings in the effective four-dimensional 
$\cx N=1$ supergravity theory in the following section.

\subsec{Effective $\cx N=1$ supergravity}
According to the above discussion  
the $\cx N=1$ superpotential appearing in the $\cx N=1$ supergravity function 
\gfun\ is 
\eqn\wwi{
W(z,\zh)= 
\sum_{\Si} \ux{N}_\Si \int_{\ga_\Si}\ux \Om ^{(3,0)}, \qquad \ga_\Si \in H_3(Z^*,\cxH)
}
if we consider the brane compactification $(Z^*,E)$ as in refs.~\refs{\LMW,\JS,\AHMM}, or, 
alternatively
\eqn\wwii{
W(z,\zh,S)=\sum_{\Si} {N}_\Si \int_{\ga_\Si}\Om^{(4,0)},\qquad \ \ga_\Si \in H_4(X^*)\ ,
}
for the F-theory compactification on the dual four-fold $X^*_{c}$.\foot{See 
also refs.~\refs{\Ltd,\PMff}\ for an early discussion of four-fold periods and superpotentials in lower dimensions.}
As discussed above, the difference between the four-fold periods and the (relative) 
three-fold periods are subleading corrections in small $g_s$; see app.~A for some details of the computation and a precise match between the periods in the example.

As for the K\"ahler potential, consider first the 
$\cx N=2$ K\"ahler potential on the base of the fibration $\cx M$, that is on 
the complex structure moduli space $\cx M_{CS}$
for the string compactification on $Z^*$ without branes. This is 
given by \ref\specg{S.~Ferrara and A.~Strominger,
``N=2 space-time supersymmetry and Calabi--Yau moduli space", in Proceedings of 
College Station Workshop (1989) 245, World Scientific;
A.~Strominger, ``Special Geometry,''
Commun.\ Math.\ Phys.\  {\bf 133}, 163 (1990); %%CITATION = CMPHA,133,163;%%
P.~Candelas and X.~de la Ossa, ``Moduli space of Calabi--Yau manifolds,''
Nucl.\ Phys.\  B {\bf 355}, 455 (1991). %%CITATION = NUPHA,B355,455;%%
}
$$
K_{CS}(z;\bb z)= -\ln Y_{CS} \ ,\qquad Y_{CS} = -i \int_{Z^*}\Om\wedge \bb \Om =
-i \sum_{\ga_\Si\in H_3(Z^*)}\, \Pi^\Si(z)\, \eta_{\Si \La}\,  \bb \Pi^\La(\bb z)\ .
$$
Here $\Si,\La =1,...,h^3(Z^*)$ and $\eta_{\Si\La}$ is the symplectic intersection matrix on $H_3(Z^*,\IZ)$, 
which represents the constant, topological metric on the space of groundstates 
in the SCFT \BCOV. Restricting the sum in eq.\wwi\ to the ``flux'' superpotential, that is
to the absolute cohomology $H_3(Z^*)$, one obtains a function
\eqn\KPcs{
\cx G_{CS} =K_{CS}(z;\bb z)+\ln W_{CS}(z)+\ln \bb W_{CS}(\bb z), \qquad 
W_{CS}(z)=\sum_{\ga_\Si\in H^3(Z^*)}N_\Si \Pi_\Si(z)\ ,}
that depends only on the closed string moduli and 
is invariant under K\"ahler transformations generated by rescalings of 
the holomorphic (3,0) form, $\Om\to e^f\Om$.

We will now give two independent arguments, that the $\cx N=1$ K\"ahler potential
on the full $\cx N=1$ deformation space $\cx M$ can be written, to leading order in $g_s$,
as
\eqn\defKP{
K(z,\zh;\bb z,\bb \zh)= -\ln Y \ ,\qquad Y =
-i\hskip-5pt\sum_{\ga_\Si \in H_3(Z^*,\cxH)}\, \ux \Pi^\Si(z,\zh)\,{\ux  \eta} _{\Si \La}\,  \ux {\bb \Pi}^\La(\bb z,\bb {\zh})\ .
}
with a pairing matrix $(\ux \eta)$ defined below.
Indeed this ansatz is a natural guess in view of the extension of the summation 
from $H^3(Z^*)$ to $H^3(Z^*,\cxH)$ in the $\cx N=1$ superpotential \wwi\ and 
defines an $\cx N=1$ supergravity function 
$\cx G(z,\zh;\bb z,\bb \zh)$ 
which is invariant under K\"ahler
transformations generated by rescalings of the relative (3,0) form 
$$\ux \Om\to e^f\ux \Om,\qquad \Pi_\Si\to e^f\ux \Pi_\Si\ . $$ 
Note that since the K\"ahler metric for the closed and open string deformations 
arises from different world-sheet topologies, the pairing $({\ux  \eta})$
on $H_3(Z^*,\cxH)$
necessarily mixes terms of different order in the string coupling $g_s$.

The first argument
comes from the results of ref.~\JL\ on the effective space-time 
action for orientifold compactifications of $D7$-branes. It has been shown
there, that the K\"ahler metric obtained by dimensional reduction of a
$D7$-brane worldvolume wrapping the orientifold plane $\cxH^\sharp$  in an orientifold 
$Z^\sharp$ is consistent, at first order in the brane
deformation, with a K\"ahler potential $K=-\ln Y_{OF}$ with
\eqn\drof{
Y_{OF}=-i \int_{Z^\sharp} (P\SS3 \ux \Om)\,  \wedge (P\SS3 \ux {\bb \Om})\,
+  \tx g \int_{\cxH^\sharp} (P\SS2 \ux \Om)\,  \wedge (P\SS2 \ux {\bb \Om})\ .
}
Here $P\SS3$ and $P\SS2$ are projection operators onto the two summands in 
eq.~\esrc, and $\tx g$ is $g_s$ times a constant. 
This is of the form \defKP\ with the pairing matrix
\eqn\anseta{
\big(\ux  \eta\big)=\pmatrix{\eta_{Z^*}&0\cr0&i\tx g\, \tx \eta_{\cxH^\sharp}}\ ,
}
where $\tx \eta_{\cxH^\sharp}$ is the (symmetric) intersection matrix on $H^2_{var}(\cxH^\sharp)$.

The second argument is obtained by computing the K\"ahler potential of the 
dual four-fold $X^*_c$, which is of a similar form as eq.\defKP\ \GMPh: 
\eqn\defKPii{
K(z,\zh,S;\bb z,\bb \zh,\bb S)= -\ln Y \ ,\qquad Y =
\sum_{\ga_\Si \in H_4(X^*)}\, \Pi^\Si(z,\zh,S)\,{\eta} _{\Si \La}\, {\bb \Pi}^\La(\bb z,\bb {\zh},\bb S)\ .
}
Here $\eta$ denotes the topological intersection matrix on $H_4(X^*_c)$ and $S$
is the afore mentioned extra modulus in the compact manifold.
An explicit computation\foot{See app.~A. for details.}  
in the weak coupling limit $\Im S \to \infty$ then shows that the four-fold K\"ahler potential can be rewritten,
to leading order in $g_s$, as the sum of two terms,  corresponding to a split \anseta\ 
with the two blocks given by the symplectic form $\eta_{Z^*}$ on 
$H_3(Z^*,\IZ)$ and $\tx \eta_{\cxH}$ 
the intersection matrix on the hypersurface $\cxH$. Thus, to leading order in $g_s$,
the F-theory result is in perfect agreement with the local orientifold result of ref.~\JL. 

The role of the
intersection matrix $\tx \eta_{\cxH}$ as defining a topological metric in the open-string sector
is natural in view of the localization of the open-string degrees of freedom 
on $\cxH$. In the quintic example, $\cxH$ describes the K3 geometry that 
captures the variation of the chain integrals with the open-string deformations, as described
below \bdrycont, 
and $\tx \eta_{\cxH}$ is simply the K3 intersection matrix $\hx \eta$ discussed below \Trel. Explicitly, the resulting K\"ahler potential for the four-fold $X^*_c$ defined by eqs.\defverts,\evert, 
reads, to leading order in small $g_s$ and in an expansion near the large complex structure point, 
%  $\phantom{\pmatrix{1\cr 1}}$
%
\eqn\KPQ{\eqalign{
K&=-\ln (\tx g^{-1}\,\tx Y),\qquad  \tx Y = \fc{5i}{6}(t-\bb t)^3+\tx g \big(-\fc{1}{6}(t_1-\bb {t_1})^4 +\fc{5}{12}(t-\bb t)^4\big)+\cx O(|t|^2)\ ,
}}
where $(t,t_1)$ are the flat coordinates of eq.\defla\ and $\tx g^{-1} = 2\,\Im S$. The second summand in 
the order $\tx g^1$ term is a correction predicted by the dual F-theory four-fold, which is not captured 
by the dimensional reduction, eq. \drof.
By mirror symmetry, eq.~\KPQ\ then represents a prediction for the 
K\"ahler metric on the deformation space of $A$-type branes on 
the quintic $Z$. It would be interesting to understand the relation of the above proposal to the
metric described by Hitchin in ref.~\ref\Hitchin{N.~J.~Hitchin,
``The moduli space of special Lagrangian submanifolds,''
Annali Scuola Sup.\ Norm.\ Pisa Sci.\ Fis.\ Mat.\  {\bf 25}, 503 (1997)
[arXiv:dg-ga/9711002].
%%CITATION = ASPFA,25,503;%%
}.

\newsec{Summary and outlook}
In this work we analyzed the deformation problem of certain families of toric 
D-branes in compact Calabi-Yau three-folds, defined along the lines of ref.\AV.
This is achieved by studying the variation of Hodge structure
as described by the periods of 
the holomorphic three-form of the Calabi-Yau manifold while keeping
track of the boundary contributions relative to a family of four-cycles describing
the B-brane geometry. We demonstrate our techniques with a specific B-brane 
configuration
in the mirror quintic. Although this geometry serves as a guiding example throughout
the paper,  we present also a general toric description of 
the generalized hypergeometric systems governing the toric brane configurations, for which our discussion applies.

We find that, similarly to the well-studied deformation problem in the pure 
closed-string sector, the notions of flatness and integrability of the Gauss-Manin
connection continue to make sense on the open-closed deformation space $\cx M$
of the family and lead to sensible results for open string enumerative invariants. Amongst others, the Gauss-Manin connection in flat coordinates displays an interesting ring structure on the infinitesimal deformations in $F^2$, which is compatible
with CFT expectations and gives evidence for the existence of an $A$-model quantum product defined by the Ooguri-Vafa invariants. Other hints in this direction are the integrability condition and the meaningful definition of the mirror map \mima\ via a flatness condition. 

For geometries with a single open string modulus the integrability conditions imply that the
relative period matrices and the Gauss-Manin connection matrices can
all be expressed in terms of functional relations involving only  
the holomorphic prepotential $\cx F$ and one additional holomorphic function $W$.\foot{This can be generalized to cases with several open-string moduli studied in \wip.}

The analyzed open-closed deformation problem can also be related to 
CFT correlators. We explained how an a priori unobstructed 
deformation problem of B-branes wrapping a holomorphic family of four-cycles 
describes an obstructed deformation problem after turning on $D5$-brane
charges. In particular this effect can be described in the CFT by 
the change of boundary 
conditions induced by non-trivial fluxes on the worldvolume of the B-brane. 
The afore mentioned holomorphic function $W$ in the
Gauss-Manin connection then turns into a superpotential encoding these
obstructions. 

By mirror symmetry our analysis carries over to the quantum integrable structure
of the obstructed deformation space of the mirror A-branes
in the mirror three-fold. For our explicit example we obtain predictions for the 
Ooguri-Vafa invariants of the open-closed deformation space that satisfy the 
expected integrality constraints and further consistency conditions.

Turning to the effective four-dimensional $\cx N=1$ supergravity theory one observes that the open-closed deformation space $\cx M$ is a fibration $\pi:\ \cx M\to \cx M_{CS}$ over the complex structure moduli and defines a K\"ahler manifold that is not of the most general form allowed by supergravity, but has a restricted ``$\cx N=1$ special geometry''. By constructing a class of dual Calabi--Yau four-folds for F-theory compactification we derived an expression for the effective $\cx N=1$ K\"ahler potential and the superpotential on $\cx M$ in terms of period integrals. 
The F-theory compactification on the dual four-fold provides a global embedding of the $B$-brane geometry on the three-fold and reduces to a local description in the decompactification/weak-coupling limit \wcl. The effective description obtained in this limit is in good agreement with the results obtained for $D7$-branes on orientifolds in the existing literature.  

However, the $B$-model results of this paper, obtained predominantly from a Hodge theoretic approach, raise also a number of
unanswered questions.
The first is about the meaning of mirror symmetry between open-closed deformation spaces in the presence of a non-trivial superpotential, which requires some sort of off-shell concept of mirror symmetry. As discussed above, a heuristic ansatz might be to define $\cx M$ first as the deformation space of an unobstructed family and then add in obstructions as a sort of perturbation, here represented by $D5$-brane charges. However, we feel that there should be a more fundamental answer to this important issue.

Another set of urgent questions concerns the $A$-model interpretation, such as a proper formulation of an $A$-model quantum ring that matches the ring structure observed on the $B$-model side and should include the Ooguri-Vafa invariants and Floer (co-)homology as essential ingredients. Similarly one would like to have a more explicit description of the target space geometry of the $A$-branes. We hope to come back to these questions in the future.

\lref\Waii{J.~Walcher,
``Evidence for Tadpole Cancellation in the Topological String,''
arXiv:0712.2775 [hep-th].}
\lref\KW{ D.~Krefl and J.~Walcher,
``Real Mirror Symmetry for One-parameter Hypersurfaces,''
JHEP {\bf 0809}, 031 (2008)
[arXiv:0805.0792 [hep-th]].}
\lref\Wanew{J.~Walcher,
``Calculations for Mirror Symmetry with D-branes,''
arXiv:0904.4905 [hep-th].
%%CITATION = ARXIV:0904.4905;%%
}
\lref\GHKK{T.~W.~Grimm, T.~W.~Ha, A.~Klemm and D.~Klevers,
``The D5-brane effective action and superpotential in N=1
compactifications,''
arXiv:0811.2996 [hep-th].}
The Hodge theoretic approach to open-string mirror symmetry is somewhat complementary to other approaches to open-string mirror symmetry in the recent literature. For compact manifolds these include the study of critical superpotentials without open moduli in refs.~\refs{\Wa,\MW,\Waii,\KW,\Wanew} and the computation of effective superpotentials in terms of matrix factorizations and related techniques \refs{\BBG,\KSi,\KSii,\Kiii,\DouglasFR,\GovindarajanUY,\AspinwallBS}. It is likely that the ``off-shell'' description of the open-closed deformation space advocated for in this paper is useful to study the interesting phenomena of phase transitions of domain walls \refs{\Wanew,\LM}, which resemble similar phase transitions in the closed-string sector, such as conifolds and flops.

\vskip2cm
\noi {\bf Acknowledgements:}\break 
We would like to thank 
Mina Aganagic,
Ilka Brunner,
Shamit Kachru,
Johanna Knapp,
Luca Martucci,
Christian R\"omelsberger,
Emanuel Scheidegger and
Johannes Walcher
for valuable discussions and comments.
The work of M.A. and P.M. is supported by the program
``Origin and Structure of the Universe'' of the German Excellence Initiative.
The work of M.H. is supported by the Deutsche Forschungsgemeinschaft.
The work of H.J. and M.S. is supported by the Stanford Institute of Theoretical Physics
and by the NSF Grant 0244728. The work of A.M. is supported by the 
Studienstiftung des deutschen Volkes. M.S. is also supported by the Mellam
Family Graduate Fellowship. 

\vskip 2cm
%%%%%%%%%%%%%%%%
%%%%%%%%%%%%%%%%
\lref\KLRY{
A.~Klemm, B.~Lian, S.~S.~Roan and S.~T.~Yau,
``Calabi-Yau fourfolds for M- and F-theory compactifications,''
Nucl.\ Phys.\  B {\bf 518}, 515 (1998)
[arXiv:hep-th/9701023].
%%CITATION = NUPHA,B518,515;%%
}
\lref\SVW{
S.~Sethi, C.~Vafa and E.~Witten,
``Constraints on low-dimensional string compactifications,''
Nucl.\ Phys.\  B {\bf 480}, 213 (1996)
[arXiv:hep-th/9606122].
%%CITATION = NUPHA,B480,213;%%
}
\vfil
\goodbreak
\lref\frob{
S.~Hosono, A.~Klemm, S.~Theisen and S.~T.~Yau,
``Mirror symmetry, mirror map and applications to complete intersection
Calabi-Yau spaces,''
Nucl.\ Phys.\  B {\bf 433}, 501 (1995)
[arXiv:hep-th/9406055];\br
%%CITATION = NUPHA,B433,501;%%
S.~Hosono, B.~H.~Lian and S.~T.~Yau,
``GKZ generalized hypergeometric systems in mirror symmetry of Calabi-Yau
hypersurfaces,''
Commun.\ Math.\ Phys.\  {\bf 182}, 535 (1996)
[arXiv:alg-geom/9511001].
%%CITATION = CMPHA,182,535;%%
}
\lref\BMft{
P.~Berglund and P.~Mayr,
``Heterotic string/F-theory duality from mirror symmetry,''
Adv.\ Theor.\ Math.\ Phys.\  {\bf 2}, 1307 (1999)
[arXiv:hep-th/9811217].
%%CITATION = 00203,2,1307;%%
}
\lref\KLRY{
A.~Klemm, B.~Lian, S.~S.~Roan and S.~T.~Yau,
``Calabi-Yau fourfolds for M- and F-theory compactifications,''
Nucl.\ Phys.\  B {\bf 518}, 515 (1998)
[arXiv:hep-th/9701023].
%%CITATION = NUPHA,B518,515;%%
}
\lref\SVW{
S.~Sethi, C.~Vafa and E.~Witten,
``Constraints on low-dimensional string compactifications,''
Nucl.\ Phys.\  B {\bf 480}, 213 (1996)
[arXiv:hep-th/9606122].
%%CITATION = NUPHA,B480,213;%%
}
\def\Pl{\tx \Pi}
\vfil
\goodbreak
\appendix{A}{$W$ and $K$ from four-folds in the quintic example}
In the following, we hand in some technical details of the computation 
of the superpotential \wwii\ and the K\"ahler potential \KPQ\ 
from the F-theory four-fold $X_c$ presented in sect.~7 and substantiate some of the
claims made there. This includes an explicit illustration
of the relation between the four-fold flux superpotential \wwii\ and the 
three-fold superpotential \wwi\ from RR and NS fluxes in the weak coupling limit
as well as the derivation of the 
K\"ahler potential \KPQ. To avoid excessive repetitions, we refer to 
refs.~\refs{\GMPh,\PMff,\KLRY} for the details on mirror symmetry for Calabi--Yau four-folds 
and to ref.~\bmff\ for the application of toric geometry to analyze geometry of
the relevant F-theory four-folds.
In the context of open-closed superpotentials the method described below has
been previously used to compute the superpotentials for several other
examples in ref.~\AHMM.

The $A$-model four-fold $X_c$ is defined by the polyhedron 
$\D$ with vertices given in \defverts, \evert. 
We consider a phase of the K\"ahler cone described by the charge vectors given
in the text, which we reproduce here for convenience:
\eqn\cvex{\vbox{\offinterlineskip\halign{
\strut # 
&\hfil~$#$~=\hfil
&\hfil~$#$ &\hfil~$#$ &\hfil~$#$ &\hfil~$#$
&\hfil~$#$ &\hfil~$#$ &\hfil~$#$ &\hfil~$#$
&\hfil~$#$ &\hfil~$#$ &\hfil~$#$ &\hfil~$#$
\cr
&\tx l^1&(&-4&0&1&1&1&1&-1&1&0)\ ,\cr
&\tx l^2&(&-1&1&0&0&0&0&1&-1&0)\ ,\cr
&\tx l^3&(&0&-2&0&0&0&0&0&1&1)\ .\cr
}}
}
The K\"ahler form is $J=\sum_a t_a J_a$, where 
$J_a$, $a=1,2,3$ denotes the basis of $H^{1,1}(X_c)$ dual to the 
Mori cone defined by \cvex. The basic topological data are the intersections
$$
K_{abcd}=\int_{X_c}J_a \wedge J_b \wedge J_c \wedge J_d\ ,$$
which can be concisely summarized in terms of the generating function\foot{Details
on the computation of the following data from toric geometry and many sample
computations for three-fold fibered four-folds can be found in \refs{\PMff,\KLRY}.}
\eqn\ffinters{\eqalign{
\cx F_4 &= \fc{1}{4!}\int_{X_c} J^4=\fc{1}{4!}\sum_{a,b,c,d} K_{\al\be\ga\de}t^\al t^\be t^\ga t^\de\cr
&=\fc{1}{4}t_1^4+\fc{5}{3}t_1^3(t_2+\fc{1}{2}t_3)+\fc{5}{2}t_1^2t_2(t_2+t_3)
+\fc{5}{3}t_1t_2^2(t_2+\fc{3}{2}t_3)
+\fc{5}{12}t_2^3(t_2+2t_3)\ .
}}
From these intersections one obtains the following topological 
invariants of $X_c$\foot{Contrary to first appearances, our choice of 
$X_c$ is not at all related to a preference for a particular soccer club.}
\eqn\topinv{\eqalign{
\chi&=\int_{X_c} c_4=1860=12\ {\rm mod}\ 24\, , \hskip2cm R_4=\int_{X_c} c_2^2=1100,\cr
R_2&=\int_{X_c} J\wedge J\wedge c_2=
90 J_1^2+110 J_2^2+ 220 J_1 J_2 + 100 J_3(J_1+J_2)\, ,\cr
R_3&=\int_{X_c} J\wedge c_3=
-330 J_1-410 J_2  -200 J_3 \, ,
}}
where $c_k$ denote the Chern classes of $X_c$. The independent Hodge numbers 
$h^{1,1}=3$, $h^{1,2}=0$, $h^{1,3}=299$ can be read off from the toric polyhedra
\refs{\PMff,\KLRY}
and fix $h^{2,2}=12+\fc{2}{3}\chi+2h^{1,2}=1252$; we refer to \SVW\ for more details and the meaning of the mod 24 condition on $\chi$.

After a linear change of coordinates, the form of the generating 
function $\cx F_4$ simplifies to  
\eqn\ffii{
\cx F_4= S \ \fc{5}{6}t^3\ + \ \big(\fc{5}{12}t^4-\fc{1}{6}t_1^4\big)\ ,
}
where  $t_3=S$ is the K\"ahler modulus of the 
$\IP^1$ base, $t_1+t_2=t$ is the modulus of the generic quintic three-fold
fiber and $t_2=\that$ will be related to the open string modulus denoted
by the same letter in sect.~4. 
Note
that the fibration structure becomes explicit in the coordinates $(S,t)$ 
and that the leading term in large $S$ contains the intersection form
on the quintic fiber $Z$. To simplify some of the following expressions, 
we use the linear combinations
$$
t'_1=t=t_1+t_2,\qquad t'_2=\that-t=-t_1,\qquad t'_3=S=t_3\ .
$$

To compute the superpotential and the K\"ahler potential we have to 
determine the integral periods of the elements of the 
vertical subspace $H^{2q}_{ver}(X_c)$ generated by wedge products 
of the elements $J_a\in H^2(X_c)$ over topological cycles in $H_{2q}(X_c,\IZ)$,
for $q=0,...,4$. Except for $q=2$, 
there is a canonical basis for $H_{2q}(X_c,\IZ)$ given
by the class of a point, the class of $X_c$, the divisors dual to the 
generators $J_a$
and the curves dual to these divisors, respectively.\foot{To be precise,
some of these basis elements may actually be integral multiples of the generators of
$H_{2q}(X_c,\IZ)$ on the hypersurface.} In an expansion 
near the large volume point $\Im t_a\to \infty\ \forall a$, the leading part
of the periods, denoted by  $\Pl_{q,.}$, 
is, up to a sign, the K\"ahler volume of the cycle as measured 
by the volume form $\fc{1}{k!}J^k$, explicitly 
\eqn\tperi{
\Pl_0=1,\qquad 
\Pl_{1,a}= t'_a,\qquad 
\Pl_{3,a}=-\fc{\p}{\p t'_a}\cx F_4,\qquad 
\Pl_{4}= \cx F_4\ .
}
As for the mid dimensional part, $k=2$, we choose basis elements defined by
intersections of the toric divisors $D_i=\{x_i=0\}$
$$
\ga_1=D_1\cap D_2,\qquad 
\ga_2=D_2\cap D_8,\qquad 
\ga_3=D_2\cap D_6\ .
$$
with intersection form and inverse 
\eqn\ifqii{
\big(\tx \eta\big)_{kl}= \ga_k\cap \ga_l=\pmatrix{-10&5&0\cr5&0&0\cr0&0&-4},\qquad 
\big(\tx \eta^{-1}\big)^{kl}=\pmatrix{0&\fc{1}{5}&0\cr\fc{1}{5}&\fc{2}{5}&0\cr0&0&-\fc{1}{4}}\ .
}
The leading parts of the $q=2$ periods are then 
\eqn\tperii{
\Pl_{2,1}= 5t'_1t'_3\, ,\qquad 
\Pl_{2,2}=\fc{5}{2}t_1'^2 \, ,\qquad 
\Pl_{2,3}=2t_2'^2 \, .\qquad 
}

\def\Tp{\Pi}\def\tp{\pi}
The subleading terms of the exact periods $\Tp_{q,.}$, correcting the 
leading parts $\Pl_{q,.}$ away from the large $\Im t$ limit,
come in two varieties. Firstly polynomial corrections in the $t_a$ of lower degree,
which involve the topological invariants \topinv.\foot{The coefficients
should be determined by a computation of the anomalous charges of wrapped D-branes,
similarly as in 
\ref\PMDc{
P.~Mayr,
``Phases of supersymmetric D-branes on K\"ahler manifolds and the McKay
correspondence,''
JHEP {\bf 0101}, 018 (2001)
[arXiv:hep-th/0010223].
%%CITATION = JHEPA,0101,018;%%
}.}
For simplicity, we ignore these terms in the following, 
as they can be interesting for other applications\foot{See e.g. 
\ref\BBCQ{
V.~Balasubramanian, P.~Berglund, J.~P.~Conlon and F.~Quevedo,
``Systematics of Moduli Stabilisation in Calabi-Yau Flux Compactifications,''
JHEP {\bf 0503}, 007 (2005)
[arXiv:hep-th/0502058].
%%CITATION = JHEPA,0503,007;%%
}.}
but do not matter for the present purpose.

Secondly, there are instanton corrections $\sim q_a=\exp{2\pi i t_a}$, 
which will be important in the match between the open string
superpotential \wwi\ and the F-theory superpotential \wwii\ claimed in sect.~7.
These instanton corrections can be computed by mirror symmetry of the 
four-folds $(X_c,X^*_c)$ as in \refs{\GMPh,\PMff,\KLRY} which can be
summarized as follows. 
To determine the 
instanton corrections, we may simply replace the leading periods $\Pl_q$ in 
eqs. \tperi,\tperii\ with the exact solutions to the Picard-Fuchs equations in 
eq.\qdiffops\ with the same leading part when rewritten in $t_a$ coordinates,
using the mirror map \mima.\foot{A more formal account of this relation 
between $A$- and $B$-models  
is elaborated on in \refs{\PMff,\KLRY}, using Frobenius method as in \frob.} 
The result for the
exact periods $\Tp_{q,k}=\Pl_{q,k}+\cx O(q_a)$ is of the form 
\font\mb=cmmib10 scaled 1050\def\Sb{\hbox{{\mb S}}}
\eqn\eperi{
\eqalign{
&\Tp_0=1,\cr 
&\Tp_{1,1}=t'_1\ ,\qquad \Tp_{1,2}=t'_2\ ,\qquad \Tp_{1,3}=\Sb\cdot 1 \ ,\cr
&\Tp_{2,1}=\Sb\cdot \,5t'_1+\tp_{2,1},\qquad \Tp_{2,2}=-\tx F_t,\qquad 
\Tp_{2,3}=-\tx W,
\cr
&\Tp_{3,1}=\Sb\cdot \, \tx F_t+\tp_{3,1}\ ,\qquad \Tp_{3,2}=\tx T\ ,\qquad 
\Tp_{3,3}=-\tx F_0\ ,\cr
&\Tp_{4}=\Sb\cdot \,\tx F_0+\tp_4\ ,
}}
where we have again indicated the fibration structure by explicitly indicating
powers in the base volume $S=t_3=t_3'$.
The other functions $\tp_{q,k}$ denote
subleading corrections whose precise form does not matter for the moment.

The leading terms in weak coupling limit $\Im S\to \infty$ are characterized by the K\"ahler
moduli $t,\that$ and the four functions 
$\tx F_t$, $\tx F_0$ and $\tx W$, $\tx T$. The first two functions 
are equal to the periods $F_t=\p_t \cx F(t)$ and 
$F_0=2\cx F(t)-t\p_t \cx F(t)$ of the quintic three-fold up to instanton corrections
in $q_S=\exp(2\pi i S)$, where $\cx F(t)$ denotes the exact instanton corrected 
prepotential of the quintic
$$
\tx F_t=F_t(t)+\cx O(q_S)=-\fc{5}{2}t^2+...,\qquad \tx 
F_0=F_0(t)+\cx O(q_S)=\fc{5}{6}t^3+...\ .
$$
Similarly, the remaining functions $\tx W$, $\tx T$ 
agree with the domain wall tension and the top period computed in sects.~3-5, 
up to instanton corrections in $S$
$$
\cx W = W +\cx O(q_S)=-2t_2'^2+...\ ,\qquad \tx T = T +\cx O(q_S)=\fc{2}{3}t_2'^3+...\ .
$$
Indeed, the function $W$ is of the form
$$
W=-2t_1^2 +\fc{1}{4\pi^2}\sum_{k=1}^\infty \sum_{n_a\geq 0} \fc{1}{k^2}\, N_{n_1,n_2}(q_1^{n_1}q_2^{n_2})^k \ ,
$$
where the instanton numbers $N_{n_1,n_2}$ and the classical term are exactly the same 
as in the superpotential $\cx T\SS1$ in eq.\tensions\ computed for the brane
compactification $(Z,L)$ (see table. 1\yyy). This remarkable correspondence
between disc instantons in a three-fold $Z$ and sphere instantons of a dual four-fold
$X$ \PM\  
is possible due to the coincidence of the multicover factor $\sim k^{-2}$
for discs in a three-fold \OV\ and spheres in a four-fold \refs{\GMPh,\PMff}.

{}From the above it 
follows that the exact instanton corrected four-fold periods on $X_c$
can be informally written as\foot{A similar relation holds in a more general
form for any three-fold fibration over $\IP^1$, see sect.~2 of \PMff.}
$$
\ux \Pi_\Si(X^*_c)= \cases{ (1,S) \times \Pi_\Si(Z^*) \cr \cr W(t,\that)}+ \ldots
$$
up to subleading corrections in $S$ denoted by the dots. Thus the closed-string superpotential of 
the four-fold, eq.\wwii,  agrees with the open-closed superpotential computed for the
brane compactification $(Z^*,E)$ in eq. \wwi, up to subleading corrections in $S$:
\eqn\wwii{\eqalign{
W(z,\zh,S)%&=\sum_{\ga_\Si \in H_4(X^*_c)} {\ux N}_\Si \int_{\ga_\Si}\Om^{(4,0)}\cr
=\sum_{\ga_\Si \in H_3(Z^*)} (N_\Si+S\ M_\Si) \int_{\ga_\Si}\ux\Om^{(3,0)} +\sum_{\ga_\Si \in H_3(Z^*,\cxH) \atop \p\ga_\Si \neq 0}
\hx N_\Si \int_{\ga_\Si}\ux \Om^{(3,0)}+\ldots\cr
}
}
The first term includes the three-fold flux superpotential on $Z^*$ 
from both RR and NS fluxes, in agreement with our identifcation of
the modulus $S$ with the type IIB string coupling.\foot{This has already been
observed earlier in a related context in ref.~\HaackDI.}
The last term describes 
the disc instanton corrected domain wall tension. This concludes
the derivation of our claims in sect.~7 concerning the superpotential. 

Eventually we can obtain the K\"ahler potential 
from \defKPii, using the topological metric
$$
\eta=\pmatrix{0&0&0&0&1\cr0&0&0&\one_3&0\cr0&0&\tx \eta^{-1}&0&0\cr
0&\one_3&0&0&0\cr1&0&0&0&0\cr} \ .
$$
It is not very illuminating to display the exact result 
and we restrict to discuss  
the leading order in an expansion for large $S$ and near large volume. Inserting
the periods \eperi\ into \defKPii, and keeping only the 
leading order terms one obtains the result in \KPQ. It is straightforward
to check, that this can be rewritten, to leading orders in $S$, in the
form \defKP\ with a pairing matrix $(\ux \eta)$ of the form \anseta\ with
blocks
$$
\eta_{Z^*}=
\pmatrix{0&0&0&1\cr0&0&1&0\cr0&-1&0&0\cr-1&0&0&0},\qquad 
\tx \eta_{D^\sharp}=\pmatrix{0&0&1\cr0&\fc{1}{4}&0\cr1&0&0}\ .
$$
To leading order the flat relative periods are given by 
$$
\tx{\ux\Pi}=\big( 1, t, \tx F_t, \tx F_0, t_1,\tx W,\tx T \big) 
=\big( 1, t, -\fc{5}{2}t^2, \fc{5}{6}t^3,t_1, -2t_1^2, -\fc{2}{3}t_1^3 \big) \ . 
$$ 
This substantiates our claim in sect.~7 concerning the K\"ahler potential.
\listrefs
\end